\documentclass{document-template}
\usepackage{amsmath}
\usepackage[version=4]{mhchem}
\usepackage{upgreek}
\usepackage{subcaption}
\usepackage[hidelinks]{hyperref}

\usepackage{xcolor}

\begin{document}

\pagestyle{fancy}

\title{Multi-modal data-driven microstructure characterization}

\maketitle

% Author: Please give full first and last names for authors and include * after the name of all corresponding authors

\author{Qi Zhang}
\author{Santiago Benito}
\author{Sebastian Weber}
\author{Markus Stricker*}

% Dedication

%\dedication{Optional dedication here. If no dedication is required, please leave blank}

% Affiliations: Please provide adacemic titles (Prof. or Dr.) for all authors where applicable, and include an institutional email address for all corresponding authors
\begin{affiliations}
Q. Z., M. S.\\
Interdisciplinary Centre for Advanced Materials Simulation, Ruhr University Bochum, Universit\"atsstr. 150, 44801 Bochum, Nordrhein-Westfalen, Germany\\
$^*$Email Address: markus.stricker@rub.de

S. B., S. W.\\
Chair of Materials Technology, Institute of Materials, Faculty of Mechanical Engineering, Ruhr University Bochum,
Universit\"atsstr. 150, 44801 Bochum, Nordrhein-Westfalen, Germany

\end{affiliations}

% Keywords: Please provide a minimum of three and a maximum of seven keywords, separated by commas

\keywords{EBSD, machine learning, clustering, iron ore reduction, latent space, dimensionality reduction}

% Abstract should be written in the present tense and impersonal style (i.e., avoid we), and be at most 200 words long
\begin{abstract}
Electron backscatter diffraction is one of the most prevalent techniques used for microstructural characterization.
In recent years, there has been an increase in the use of data-driven methods to analyze raw Kikuchi patterns.
However, most of these require user input and the interpretation of the data-derived features is often challenging and subject to \textit{informed interpretation}.
By using a combination of principal component analysis, constrained non-negative matrix factorization, and a variational autoencoder along with information-theoretical considerations on a multimodal dataset, it is shown that a) automated decision on method-specific hyperparameters, here the number of components in principal component analysis, the number of components for constrained non-negative matrix factorization, and the selection of reference constraints; and b) latent space features can be mapped to physically-meaningful quantities.
In addition, the recommended region-of-interest (ROI) size for optimal model performance is approximated automatically to be twice the characteristic grain size based on information content of the dataset.
Implemented in a workflow, this allows for a transferable, dataset-specific autonomous data-driven phase and grain segmentation including grain boundary detection and the analysis of very-small-angle intra-grain variations to complement conventional electron backscatter analysis.
\end{abstract}

\section{Introduction}

Material behavior is ultimately rooted in atomic interactions, however, the term \textit{More Is Different}~\cite{Anderson1972} probably describes best the emerging behavior with many interacting atoms we typically call `a microstructure'.
Hence, it is the characterization of microstructure at above-atomic level which is typically targeted in materials characterization.
The microstructure of a material largely determines its properties by giving rise to a dominating length scale.
This could be the average grain size~\cite{Hall1951,Petch1953} or the dislocation density~\cite{Taylor1934,Taylor1934a} or a $\gamma$/$\gamma'$ microstructure in superalloys~\cite{Reed2008} dominating the mechanical response of a material.

One of the most widely employed techniques for structural characterization at the nano- and micro-scale is Electron Backscatter Diffraction (EBSD).
But even adding up all characterizations done so far, the material volume which has been covered can be estimated as several tens of $\upmu$m$^3$:
Assume that an EBSD scan covers approximately $V= \pi r^2 d$ per measurement, where $r$ is half the beam diameter (typically $1\,\upmu$m) and $d$ the interaction depth (typically $100$-$500\,$nm).
Furthermore, assuming that a typically mapped area covers $100\,\upmu$m by $100\,\upmu$m, $V \approx 3 \cdot 10^{-10}\,$m$^3$ and approximately 14412 scientific articles reporting EBSD results with an estimated two EBSD maps per each results in 28824 maps, resulting in $\approx 10\times10^{-6}\,$m$^3$, ($10\,\upmu$m$^3$).
The main information extracted from each pattern typically only comprises the local structure and orientation and then derivatives of that like average grain size, precipitate size, orientation distributions, etc.

However, recent advances show that more than just structural and orientation information can be extracted from raw Kikuchi patterns for a deeper characterization than conventional EBSD analysis.
Correlative microscopy with a correction scheme allows to resolve sub 1$^{\circ}$ misorientations~\cite{Thome2019}.
Unsupervised machine learning approaches have been used for phase segmentation~\cite{McAuliffe2020}.
Constrained non-negative factorization for sub-EBSD resolution segmentation~\cite{chauniyalEmployingConstrainedNonnegative2024b}, and the extraction of latent space features based on variational autoencoders hints at the possibility to even resolve heterogeneities in microstructures like defects~\cite{calvatLearningMetalMicrostructural2025b}.
These examples are part of a surge of data-driven approaches for microstructural characterization in recent years~\cite{Holm2020}.

However, data-driven workflows such as these typically require manual input from an expert and, therefore, lack transferability to other microstructure datasets largely due to the (very) heterogeneous nature of possibilities of microstructures.
Constrained non-negative matrix factorization~\cite{chauniyalEmployingConstrainedNonnegative2024b} for instance, requires the choice of a reference region of interest to constrain the algorithm.
Or the resulting latent space features of the encoded microstructure are not automatically labeled but subject to expert interpretation and remain specific to the dataset from which they were extracted~\cite{calvatLearningMetalMicrostructural2025b}.

Here, we build upon the approach of constrained non-negative matrix factorization for the segmentation of microstructures below the level of the original resolution~\cite{chauniyalEmployingConstrainedNonnegative2024b}.
We present an end-to-end data-driven workflow that does not require any user input by a combination of methods comprising dimensionality reduction, clustering, and constrained non-negative factorization.
Our example microstructure is a partially-reduced iron ore pellet exhibiting a rich microstructure of different phases, phase boundaries, grain boundaries, pores, as well as intragranular orientation variations, namely with structural and chemical heterogeneity.
Our proposed workflow creates multiple representations of individual regions' electron backscatter patterns and we use these at different stages of the workflow for decision tasks related to hyperparameters.
Such a workflow is exemplary for how a combination of data-driven approaches \textit{together} can be implemented for a transferable and self-consistent data-driven characterization workflow for complex microstructures.

\section{Methods}
% \MS{Todo(Qi): check with Santiago}
\subsection{Data Acquisition}
Our sample is a commercial direct-reduced hematite pellet composed of 0.36 wt\% \ce{FeO}, 1.06 wt\% SiO$_2$, 0.40 wt\% Al$_2$O$_3$, 0.73 wt\% CaO, 0.57 wt\% MgO,  0.19 wt\% TiO$_2$, 0.23 wt\% V, 0.10 wt\% Mn, with P,  S, Na, and K as the minors, and \ce{Fe2O3} for the balance.
The central portion (approximately $3$–$4$\,mm thick) was removed, and the remaining half was used for subsequent characterization.
The multi-stage reduction of iron ores can be roughly assumed as phase transformations from a high oxidation state \ce{Fe2O3} to a low oxidation state (\ce{Fe3O4} and \ce{Fe1-xO}) and final product (\(\alpha\)-Fe).
Hematite exhibits a trigonal structure, represented by a hexagonal prism unit cell while the subsequently-formed magnetite (\ce{Fe3O4}), w\"ustite (\ce{Fe1-xO}), and \(\alpha\)-Fe all possess cubic symmetry.
For a polycrystalline porous industrial pellet, it is assumed that the pores inherited from the sintering stage further develop during the reduction process, and iron initially nucleates on the free surfaces, exhibiting a phase gradient along the pellet radius~\cite{maHydrogenbasedDirectReduction2022}.
A cell-like morphology develops through formation of a percolating pore network for single crystal hematite as demonstrated in~\cite{ratzkerElucidatingMicrostructureEvolution2025} fundamentally.

All Kikuchi patterns were captured in a MIRA 3 scanning electron microscope (SEM) by Tescan Orsay Holding, equipped with a field emission cathode. 
The sample preparation involved mechanical grinding of the newly exposed cross-sectional surfaces using SiC abrasive papers from 800 up to 4000 grit, followed by polishing with 3\,\(\upmu\)m and 1\,\(\upmu\)m diamond suspensions.  
The final step consisted of polishing with an oxide polishing suspension (OPS) for $5$-$10$\,min to remove the remaining surface deformation and obtain a surface quality suitable for high-quality Kikuchi patterns, after which the specimen surface was briefly cleaned with ethanol prior to Electron backscatter diffraction (EBSD)/ Energy-dispersive X-ray Spectrometry (EDS) acquisition.

The EBSD dataset was obtained with a Nordlys Nano Detector coupled with the AZtec software package by Oxford Instruments.
The specimen was tilted 70$^{\circ}$ and the following microscope settings were employed: acceleration voltage of 15\,kV, working distance of 17\,mm, beam diameter of 72\,nm.
The measurement covered an area of $13 \times 6\,\upmu$m$^2$ ($495 \times 266$ scan points) of the hematite pellet resulting in a step size of 0.0263\,$\upmu$m.
All electron backscatter Kikuchi patterns (EBSPs) were collected with an acquisition time of 0.0448\,s and $2\times2$\,pixels binning in the CCD sensor, ultimately resulting in an image 672 by 512 pixels.
These were saved as TIFF files for each individual sample point.
In addition, EDS data was acquired simultaneously with the EBSD measurements under the same scanning conditions with an Xmax 50 detector by Oxford Instruments.
%Elemental maps were acquired for 13 elements across each scan point, resulting in a quantitative dataset of elemental concentrations spatially corresponding to the EBSD grid.

Besides capturing, encoding, and saving the EBSPs, we also employed the AZtec software by Oxford Instruments to perform some tasks that conventionally belong to the pre- and post-processing of the data, including some quantitative analysis.
For instance, using the Inorganic Crystal Structure Database (ICSD) entries for hematite (ICSD 130951), magnenite (ICSD 5247), w\"ustite (ICSD 76639), and $\alpha$-Fe (ICSD 52258), Hough-transform indexing of the EBSPs was carried out.
Further, each EBSP was assigned a value of band contrast (BC) as an indicator of its pattern quality to create a BC grid.
The EDS spectra of each recorded position were quantitatively analyzed in AZtec to produce elemental concentration maps spatially corresponding to the EBSD grid. The following elements were considered: Fe, O, Al, Ca, K, Mg, Mn, Na, P, S, Si, Ti, V.
The indexed EBSD data was exported as a CHANNEL5 project, which was then imported into the MATLAB Toolbox MTEX~\cite{HielscherSchaeben2008}.
MTEX was employed for denoising, data filling, and grain reconstruction to obtain a refined dataset.
The conventional analysis results are presented in Section~\ref{ssec:exp_analysis}.

\subsection{Data-driven approaches}
\label{ssec:method_data_driven}

\begin{figure}[htp!]
  \centering
  \includegraphics[width=\linewidth]{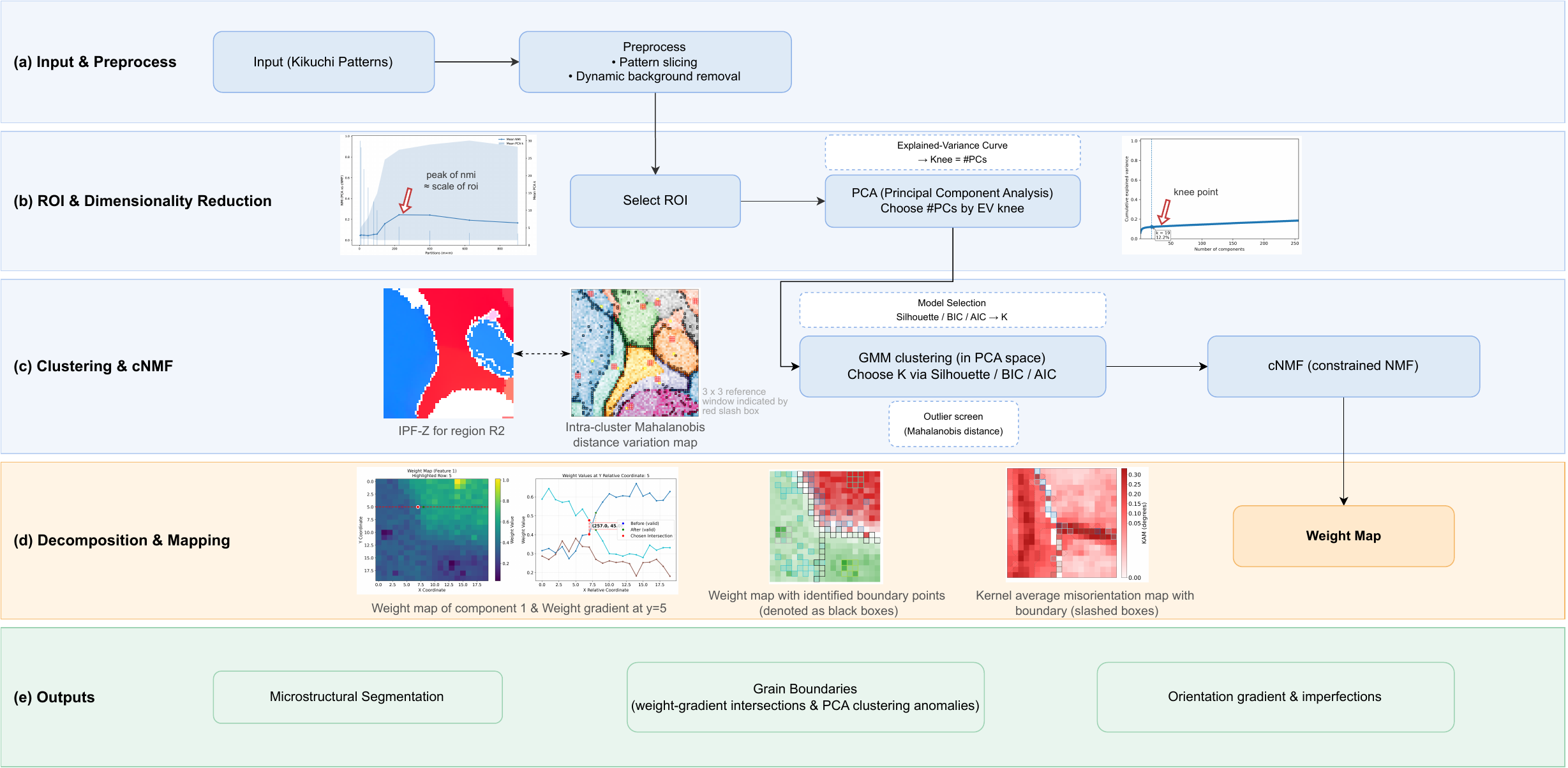}
  \caption{Workflow overview with indication of which methods are used: (a) Preprocessing, (b) Selection of the region of interest (ROI), (c) clustering and constrained non-negative matrix factorization, (d) Kikuchi pattern decomposition and segmentation, and (e) characterization outputs.}
  \label{fig:workflow}
\end{figure}

Our data-driven approach to microstructure characterization can broadly be divided into five parts as shown in Figure~\ref{fig:workflow} (a)-(e).
We start (a) with preprocessing the raw data
including dynamic background removal.
Then in (b) we decide on a discretization of the whole measurement area to be subjected to constrained non-negative matrix factorization and decide on the dimensionality of the reduced representation of Kikuchi patterns in principle component space.
In microstructural terms, this means we establish the microstructural length scale defined as the ROI scale that maximizes the normalized mutual information (NMI) between clustering results over PCA- and constrained-non negative matrix factorization (cNMF) representations which correlates with a dominating length scale like the average feature size of grain/phase extent in our case.
The next step (c) comprises the clustering of patterns represented in PCA space using a Gaussian Mixture Model (GMM) to automatically choose the number of clusters $K$ based on information criteria.
This allows the representation of Kikuchi patterns in weight space of the constrained non-negative matrix factorization.
These weights can then be used for weight maps, i.e. segmentation of the microstructure in (d), and allows further analysis of microstructural features.
Finally, we are able to identify grain and phase boundaries based on weight-gradient intersection and `anomalies' in PCA space clustering and can visualize intra-grain orientation gradients and other defects.
Following this overview, we detail all used methods.

%\MS{@Qi: What is ``pattern slicing''?}\QZ{Here, pattern slicing refers to cropping each pattern to a fixed central region by discarding the outer edge pixels that are prone to detector-edge artefacts and distortions}

\subsubsection{Principal Component Analysis and Gaussian Mixture Model clustering}
For the preprocessing of Kikuchi signals we use \texttt{kikuchipy} \cite{Anes2020_kikuchipy} prior to data-driven analysis.
Raw patterns are cropped from $512 \times 672$ pixels to $400 \times 500$ pixels to keep the central field of view and exclude the edge artefacts.
Band intensities are linearly rescaled to $0$-$1$ for signal normalization following a high-pass-style correction and dynamic background removal in the frequency domain.

As illustrated in Figure~\ref{fig:workflow}, the preprocessed signals are then flattened into one-dimensional vectors and analyzed applying Principal Component Analysis (PCA) or Incremental PCA (IPCA) depending on the size of ROI (region of interest).
PCA is a linear dimensionality reduction method that discerns a set of orthogonal projections \(W_k = [w_1, \dots, w_k]\) by maximizing the data variance as formulated by 

\begin{equation}
\max_{W_k^\top W_k = I_k} \mathrm{tr}(W_k^\top S W_k), \quad 
S = \frac{1}{n-1}\tilde{X}^\top \tilde{X},    
\label{eq:pca}
\end{equation}

where \(S\) is the covariance matrix of the centered data \(\tilde{X}\).
When the total number of patterns \(n\) is too large to fit in memory of a typical workstation, IPCA is used to perform the same decomposition iteratively by processing data in small batches.
Each batch updates the global mean and the leading eigenvectors without storing all samples at once, thus maintaining a memory footprint proportional only to the number of retained components \(k\).  
Mathematically, IPCA approximates the same PCA subspace by computing local singular value decompositions (SVDs) on batches and merging the results incrementally which enables efficient analysis of datasets containing tens or hundreds of thousands of Kikuchi patterns.

\begin{figure}
  \centering
  \includegraphics[width=0.6\linewidth]{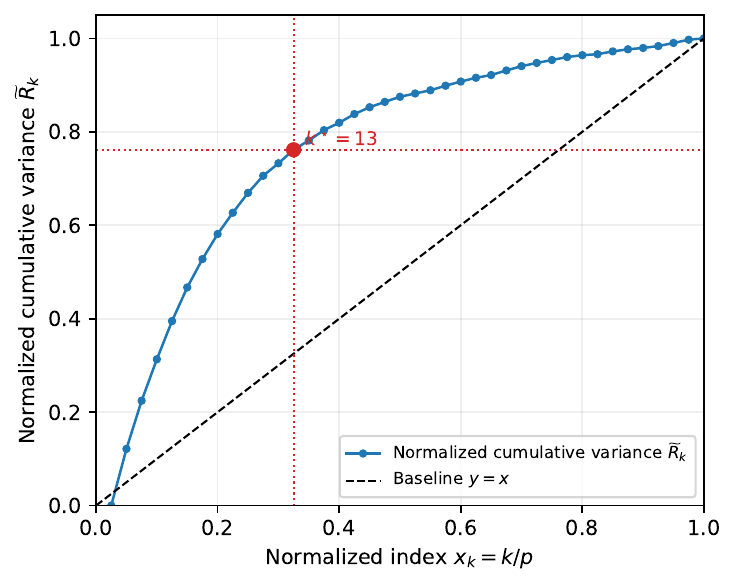}
  \caption{Normalized cumulative variance as a function of the normalized index of the number of principal components. Red marker shows the detected knee-point.}
  \label{fig:pca_knee_demo}
\end{figure}

To determine the optimal number of principal components \(k^\star\), an automatic knee-point detection algorithm is used based on the cumulative explained variance curve \(R_k\).
Intuitively, the ``knee'' marks the point beyond which adding more components yields little additional variance, i.e. the transition from signal-dominated to noise-dominated components \cite{BoehmkeGreenwell2019}.
The algorithm is visualized in Figure~\ref{fig:pca_knee_demo}:
Let \(x_k = k/p\) be a normalized component index (ranging from 0 to 1), and \(R_k\) the cumulative explained variance.
\(R_k\) is normalized to \([0, 1]\):

\begin{equation}
\widetilde{R}_k = \frac{R_k - R_1}{R_p - R_1 + \varepsilon},    
\end{equation}

and define the gap function:
\begin{equation}
g_k = \widetilde{R}_k - x_k.    
\end{equation}

The optimal component number is chosen as the index with the largest gap:
\begin{equation}
k^\star = \arg\max_k g_k,    
\end{equation}

subject to a minimum component count \(k_{\min}\) (typically 2).

It is conceptually similar to the Kneedle algorithm~\cite{Zhang2021DPZ}, and automatically finds the point where the cumulative variance curve bends away from the diagonal baseline (the line \(y=x\)).

The reduced PCA feature matrix is further analyzed using a GMM to identify statistically distinct groups of Kikuchi patterns within the dataset.
Many recent studies have successfully employed various clustering methods to segment or identify microstructural features in complex materials systems.  
For instance, Parish~\cite{parishClusterAnalysisCombined2022} proposed a data-analytics-based framework combining EDS and EBSD information to distinguish chemically distinct phases with crystallographically similar structures using singular value decomposition (SVD) followed by K-means and Fuzzy C-means clustering.
Similarly, Kim \textit{et al.}~\cite{kimUnsupervisedSegmentationMicrostructural2022} developed an unsupervised machine learning approach integrating feature extraction with a Bayesian Gaussian mixture model to segment low-carbon steel microstructures directly from optical microscopy images.  
In another example, Vincent \textit{et al.}~\cite{vincentDataClusterAnalysis2023} used a Gaussian mixture model to identify feature clusters within Raman spectra, enabling high-throughput classification of twisted bilayer graphene samples after dimensionality reduction of the spectral data.  

Unlike hard clustering methods such as $k$-means, GMM provides a \textit{probabilistic} description of the data distribution by modeling it as a mixture of \(K\) multivariate Gaussian components:
\begin{equation}
p(x) = \sum_{i=1}^{K} \pi_i\, \mathcal{N}(x \mid \mu_i, \Sigma_i),    
\end{equation}

where \(\pi_i\) denotes the mixture weight (\(\sum_i \pi_i = 1\)), and \(\mu_i, \Sigma_i\) are the mean and covariance of each cluster.  
This probabilistic framework allows GMM to capture anisotropic cluster shapes and overlapping distributions, which are appropriate to segment the reduced PCA representations of Kikuchi patterns.

GMM parameters are estimated using the Expectation–Maximization (EM) algorithm, which iteratively maximizes the likelihood of the data under the mixture assumption.  
To avoid manual specification of the cluster number, the optimal component count \(K^\star\) is determined automatically through a hybrid model selection scheme.  
Specifically, candidate models with \(K \in [K_{\min}, K_{\max}]\) are evaluated using three complementary metrics: the Akaike Information Criterion (AIC)~\cite{Akaike1974AIC}, the Bayesian Information Criterion (BIC)~\cite{Schwarz1978BIC} and the Silhouette Coefficient (SC)~\cite{Rousseeuw1987Silhouettes}.
These three metrics are typically used to determine the best number of clusters in data-driven material characterization.
For instance, Liu \textit{et al.}~\cite{liuDatadrivenCompositionDesign2024} employed AIC and BIC to determine the optimal number of clusters when constructing a GMM for composition–property mapping.
Similarly, Muir \textit{et al.}~\cite{muirDamageMechanismIdentification2021} summarized AIC, BIC, and SC as three major heuristic criteria for identifying the correct number of clusters corresponding to distinct damage modes in acoustic emission signals.  

In our case, we compute a composite score by assigning weights to these criteria,

\begin{equation}
S(K) = w_1\!\left[-\mathrm{AIC}(K)\right] + w_2\!\left[-\mathrm{BIC}(K)\right] + w_3\,\mathrm{SC}(K),
\end{equation}

where \(w_1, w_2, w_3\) represent adjustable weights reflecting the relative importance of model parsimony and geometric separability.  
The cluster number corresponding to the maximum \(S(K)\) is selected as the optimal configuration:

\begin{equation}
K^\star = \arg\max_K S(K).    
\end{equation}

This hybrid criterion balances statistical goodness-of-fit with physical interpretability and ensures that the clustering process remains fully automated, data-driven, and robust against overfitting or under-segmentation.

\begin{figure}[htp!]
  \centering
  \includegraphics[width=0.6\linewidth]{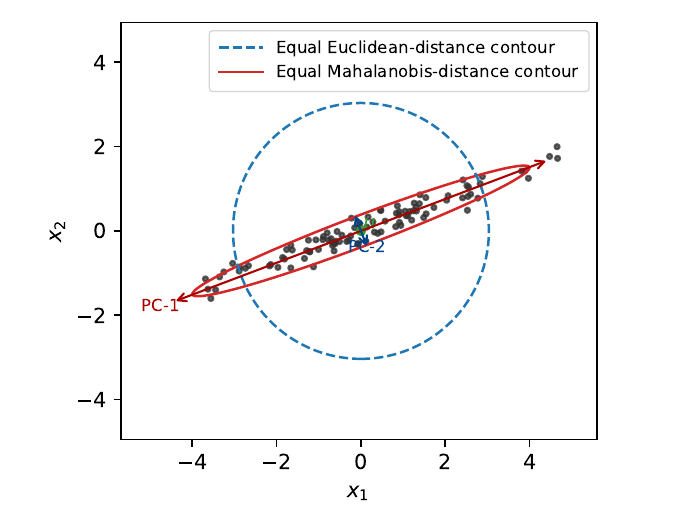}
  \caption{Mahalanobis vs. Euclidean contours (equal distance to centroid) aligned with principle components 1 and 2.}
  \label{fig:euclid_vs_mahalanobis}
\end{figure}

In addition, we define a new concept called `anomaly'. 
Anomalies are outlier patterns that statistically significantly deviate from the principal component space of each cluster.
The Mahalanobis distance (MD) metric (\(D_M\))~\cite{DeMaesschalck2000Mahalanobis} is chosen as a statistical indicator to filter the anomalies.
MD measures how far a point \(x\) lies from the multivariate mean of a distribution while accounting for the data covariance structure in the principal component space:

\begin{equation}
D_M^2(x) = (x - \boldsymbol{\mu})^\top \boldsymbol{\Sigma}^{-1} (x - \boldsymbol{\mu}),    
\end{equation}

where \(\boldsymbol{\mu}\) and \(\boldsymbol{\Sigma}\) denote the mean vector and covariance matrix of the PCA features within a given cluster, respectively.
Figure~\ref{fig:euclid_vs_mahalanobis} shows how the MD corresponds to measuring distances in a \emph{covariance-scaled} principal component space: directions with large variance are ``wider'' and penalized less, whereas directions with small variance are ``narrower'' and penalized more:
MD scales the feature space according to the cluster covariance.
This is in contrast to the Euclidean distance (ED) which assumes isotropic variance and uncorrelated dimensions~\cite{brownMahalanobisDistanceBased2022}.

Our concept of anomalies enables a consistent distance evaluation even when the principal components have unequal variances, which is often the case for data derived from complex, high-dimensional Kikuchi representations.
Notably, MD-based anomaly detection has also been widely and effectively used in structural health monitoring, where covariance-aware distance measures help identify abnormal vibration or response patterns under environmental and operational variability~\cite{sarmadiNovelAnomalyDetection2020}.

Additionally, the points exhibiting a relatively lower Mahalanobis distance within each cluster, i.e. those located closest to the cluster centroid in the covariance-weighted feature space can be regarded as the statistically most representative samples of that cluster, which subsequently are used as reference constraints in cNMF.
This allows the decision of the `constraints' for cNMF to be fully automatic and specific to the data distribution and requires no user input.

For each cluster, the squared Mahalanobis distances of all patterns are computed and compared against a threshold derived from the chi-square (\(\chi^2\)) distribution with \(d\) degrees of freedom, where \(d\) is the number of retained PCA components:

\begin{equation}
D_M^2(x) > \chi^2_{0.95}(d).    
\end{equation}

Points outside the 95\% confidence ellipses are classified as anomalies.
This criterion effectively identifies points that have less than $5$\,\% probability of belonging to the statistical population defined by their assigned cluster.

In our context, anomalies coincide with grain or phase boundaries, local heterogeneities in phase transition fronts, or intra-grain defective regions that exhibit distinct diffraction features from their surrounding patterns as will be shown later.
In any case, patterns classified as anomalies statistically significantly deviate from their naive assignment to a cluster and represent `interesting points' of a microstructure that warrant further analysis.

\subsubsection{Constrained Non-Negative Matrix Factorization}
To further decompose the Kikuchi pattern into physically-interpretable features, cNMF is used.
It is a variant of the classical non-negative matrix factorization (NMF) algorithm~\cite{maffettoneConstrainedNonnegativeMatrix2021}, which factorizes the observation matrix \( X \in \mathbf{R}_{+}^{n \times m} \) into two non-negative matrices:

\begin{equation}
X \approx W H,    
\end{equation}

where \(W \in \mathbf{R}_{+}^{n \times r}\) represents the weight matrix and \(H \in \mathbf{R}_{+}^{r \times m}\) the component (basis) matrix~\cite{maffettoneConstrainedNonnegativeMatrix2021}.  
In our study, like in~\cite{chauniyalEmployingConstrainedNonnegative2024b}, each row vector of \(X\) corresponds to a flattened Kikuchi pattern (EBSP), and \(H\) consists of the selected reference EBSPs that serve as constraints during the factorization process. 
Accordingly, the weight matrix \(W\) quantifies the content or contribution of each constrained reference component in each EBSP, allowing the local microstructural composition to be represented quantitatively.

Following the approach proposed by Chauniyal \textit{et al.}~\cite{chauniyalEmployingConstrainedNonnegative2024b}, cNMF is robust in identifying relative grain misorientations, sub-grain features, and sub-EBSD resolution identification of grain and phase boundaries.
Their study demonstrated that cNMF can detect grain boundary regions with angular deviations of less than $1^{\circ}$, aligning closely with reference orientations obtained from computationally much more demanding rotation-vector-baseline EBSD (RVB-EBSD) analysis~\cite{Thome2019}.  
However, unlike~\cite{chauniyalEmployingConstrainedNonnegative2024b}, where the constrained components were manually selected from grain interiors, our study introduces a fully automated, data-driven strategy for reference component selection based on the aforementioned Mahalanobis distance.

Specifically, for each cluster obtained in the PCA space, the ten most central samples (with the smallest Mahalanobis distance to the centroid) are identified.  
Around each of these candidates, a \(3\times3\) physical-space window is extracted, and its statistical compactness is evaluated by computing the mean and variance of the Mahalanobis distances within the chosen $3\times 3$ window of candidate reference regions:

\begin{equation}
\text{Metric} = w_1 \cdot \overline{D}_M + w_2 \cdot \mathrm{Var}(D_M),    
\end{equation}

where \(w_1\) and \(w_2\) can be weighting factors but are here set to unity.
The window exhibiting the minimum of this metric is selected as the optimal \textit{reference window} for that cluster, representing the most homogeneous and `statistically stable' local region.
The corresponding averaged and dynamically background-corrected Kikuchi patterns within each selected window are then processed and flattened to form the component constraints \(H\) for the subsequent cNMF decomposition.

These automatically-determined reference components capture the \textit{prototypical} diffraction features of each phase or substructure class without user input.
Inferred from the findings of Chauniyal \textit{et al.}~\cite{chauniyalEmployingConstrainedNonnegative2024b}, the spatial variation of the resulting cNMF weight maps reflects deviations from the reference patterns and thus highlights microstructural transition regions such as grain or phase boundaries.  
In the present implementation, the boundary regions are quantitatively defined as the intersection points along the sample's \(x\)- or \(y\)-direction where the cNMF weight contributions from neighboring components intersect.
A detailed interpretation of these weight-field variations and their correlation with physical microstructural features is presented in Section~\ref{sec:results} and~\ref{sec:discussion}.

\subsubsection{Variational Autoencoder}
While the cNMF workflow focuses primarily on local, region-specific variations within a defined prior region of interest (the manually or statistically determined constraints), it remains inherently affected by the expressivity of the predefined components, namely the resulting decomposition can only capture structural variations that are already represented by chosen reference components.
In contrast, a Variational Autoencoder (VAE) offers a complementary, data-driven approach capable of learning smooth latent-space representations that capture both observed and unseen structural features without the need for explicit prior constraints~\cite{liuGenerativeArtificialIntelligence2023}.  

A recent study has demonstrated the potential of VAE-based frameworks for microstructure informatics, i.e. novel ways to extract more information from the same data.
Calvat \textit{et al.}~\cite{calvatLearningMetalMicrostructural2025b} developed a data-reduced representation of metal microstructures and constructed spatial mappings of the latent space features to capture the heterogeneity of diffraction-based signals.  

\begin{figure}[htp!]
  \centering
  \includegraphics[width=\linewidth]{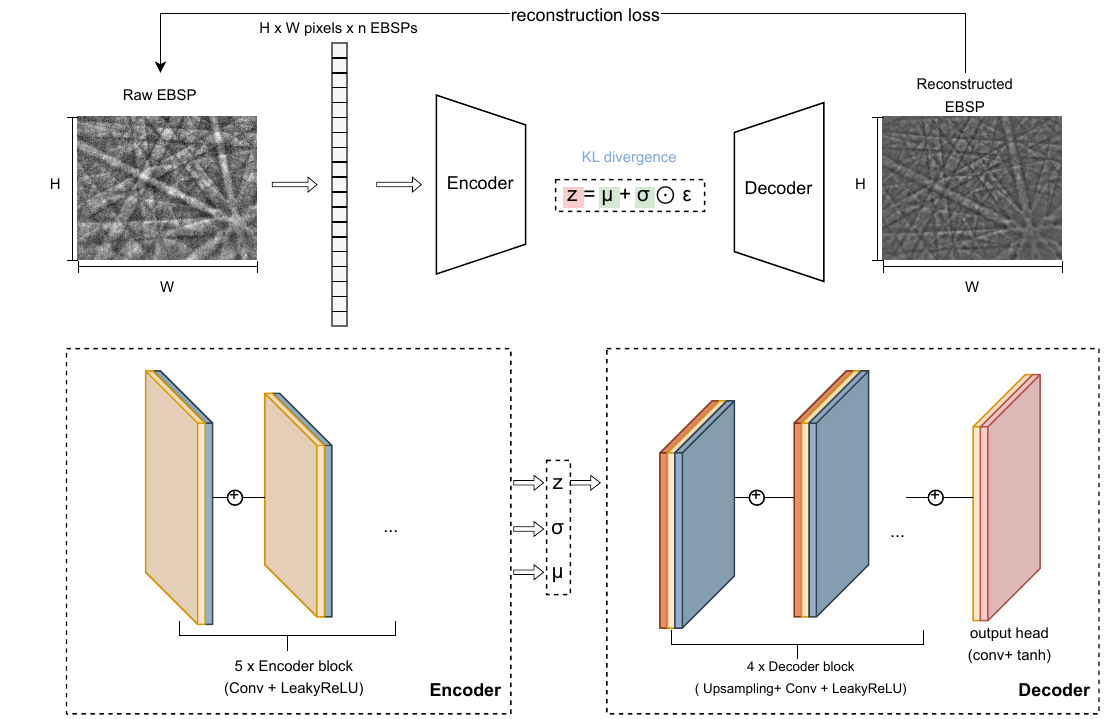}
  \caption{Variational autoencoder architecture framework for Kikuchi pattern latent space creation.}
  \label{fig:vae}
\end{figure}

We also implement a convolutional VAE architecture as shown in Figure~\ref{fig:vae} to achieve similar dimensionality reduction with latent space microstructural features.
The encoder network comprises five convolutional blocks, each consisting of a $3\times3$ convolution (stride 2, padding 1) followed by a LeakyReLU activation (\(\alpha=0.2\)), with progressively increasing channel depth from 32 to 512.

The encoded feature maps are flattened and projected into two vectors, \(\boldsymbol{\mu}\) and \(\log \boldsymbol{\sigma}^2\), representing the mean and logarithmic variance of the latent distribution (\(z \in \mathbf{R}^{512}\)):

\begin{equation}
z = \boldsymbol{\mu} + \exp\left(0.5 \cdot \log \boldsymbol{\sigma}^2\right) \odot \boldsymbol{\varepsilon}, \quad \boldsymbol{\varepsilon} \sim \mathcal{N}(0, I).    
\end{equation}

The decoder network mirrors the encoder in reverse, beginning with a fully connected projection reshaped to \((C, H/32, W/32)\), followed by four upsampling blocks (nearest or bilinear interpolation) and $3\times3$ convolutions with LeakyReLU activations.  
The final reconstruction layer employs a $3\times3$ convolution followed by a $\tanh$ activation to generate normalized output patterns.
The VAE is trained to minimize a combined reconstruction and regularization loss:

\begin{equation}
\mathcal{L} = \mathrm{MSE}(x, \hat{x}) + \beta \, \mathrm{KL}\!\left(q_{\phi}(z|x) \,||\, \mathcal{N}(0, I)\right),    
\end{equation}

where \(\mathrm{MSE}(x, \hat{x})\) quantifies the reconstruction fidelity, and the Kullback-Leibler-divergence (KL) term regularizes the latent distribution toward an isotropic Gaussian prior.  
The weighting coefficient \(\beta\) (\(=\) KL weight) controls the trade-off between reconstruction accuracy and latent smoothness.  
This model enables a compact yet physically potentially meaningful embedding of diffraction patterns.
Furthermore, we correlate the resulting learned latent features with EDS elemental distributions and band contrast to assess their physical interpretability, and the comparative strengths and limitations of the VAE and cNMF workflows are systematically evaluated.

\section{Results}
\label{sec:results}

We present our results in two parts:
First, we present a standard EBSD-based analysis of the partially-reduced iron ore pellet including phase segmentation, inverse pole figure, and grain size distribution analysis in Section~\ref{ssec:exp_analysis}.
Second, in Section~\ref{ssec:data_driven_characterization}, we present our self-consistent, unsupervised data-driven analysis using the methods described in Section~\ref{ssec:method_data_driven} on the same dataset to highlight that a) data-driven methods do provide additional insights and b) a largely unsupervised workflow can be implemented without user input.

\subsection{Experimental Analysis}
\label{ssec:exp_analysis}
\begin{figure}[htp!]
  \centering
  \includegraphics[width=.6\linewidth]{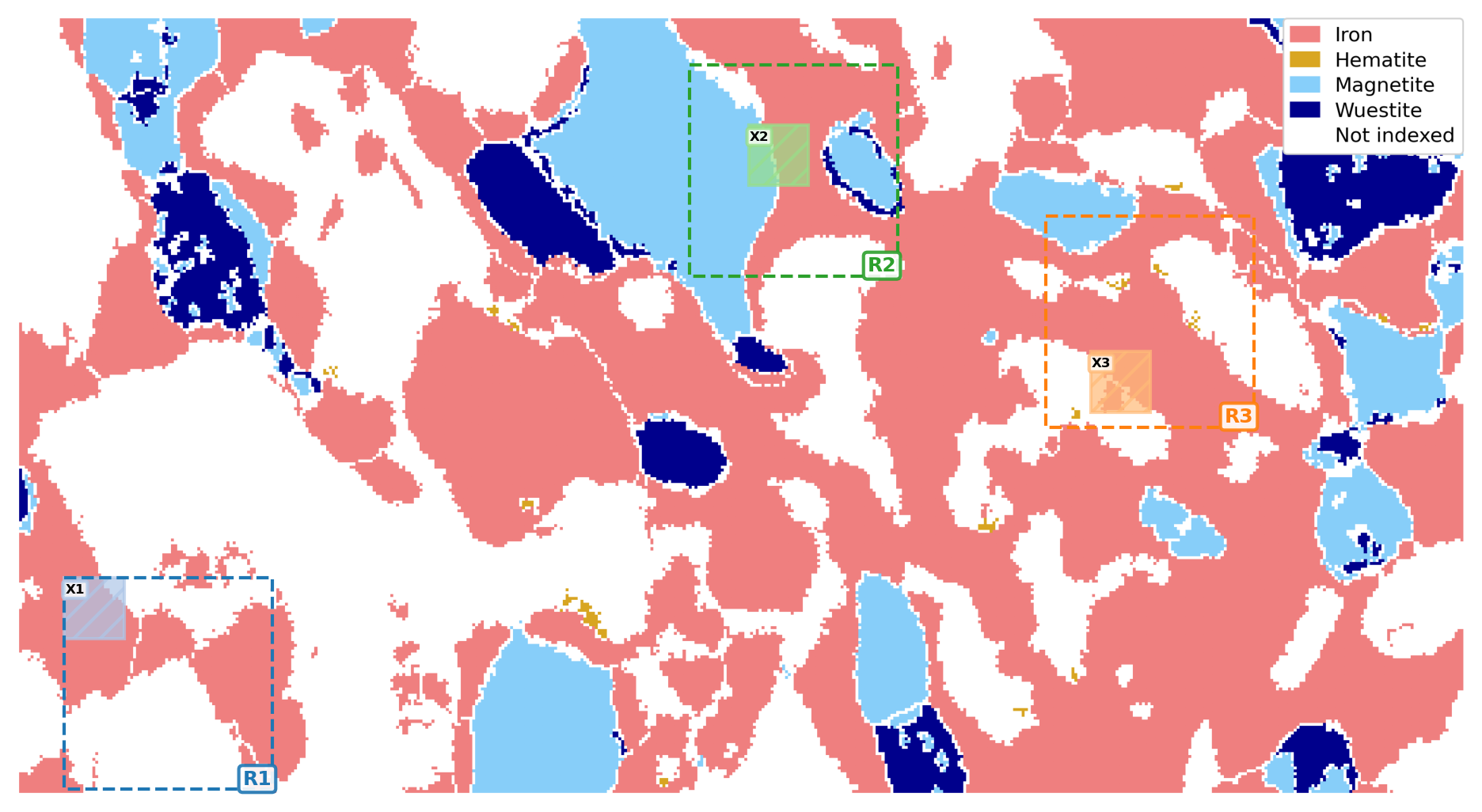}
  \caption{Conventional analysis phase map with selected ROIs (R1-R3; X1-X3; X regions within R regions)}
  \label{fig:phase_map_with_rois}
\end{figure}

For the experimental EBSD data analysis and initial indexing we use the Oxford AZtec EBSD system and the accompanying software.
Subsequently, the raw EBSD datasets are post-processed using the MTEX toolbox~\cite{HielscherSchaeben2008} in MATLAB to perform denoising, smoothing, and interpolation of missing data, resulting in a refined, continuous phase map shown in Figure~\ref{fig:phase_map_with_rois}.

The processed phase map reveals three primary phases present in the reduced pellet as expected from a partially-reduced iron ore pellet: Magnetite (\ce{Fe3O4}), w\"ustite (\ce{FeO}), and \(\alpha\)-iron (\ce{Fe}).
Three regions of interest (ROIs), denoted as R1, R2, and R3, are selected from the phase map, each covering an area of approximately \(1.8 \times 1.8\,\upmu\mathrm{m}^2\) corresponding to \(70 \times 70\) scan points (samples).
Within each ROI, a smaller subregion (X1, X2, and X3, respectively) is defined to provide a high-resolution reference for subsequent data-driven cNMF analysis.  
Each subregion X covers an area of approximately \(0.5 \times 0.5\,\upmu\mathrm{m}^2\) (\(20 \times 20\) scan points).
Additionally, a significant fraction of pores is observed within the microstructure, which can be attributed both to those inherited from the sintering stage and to newly generated pores formed during the reduction process due to oxygen transport and water removal.  

\begin{figure}[htp!]
  \centering
  \includegraphics[width=.6\textwidth]{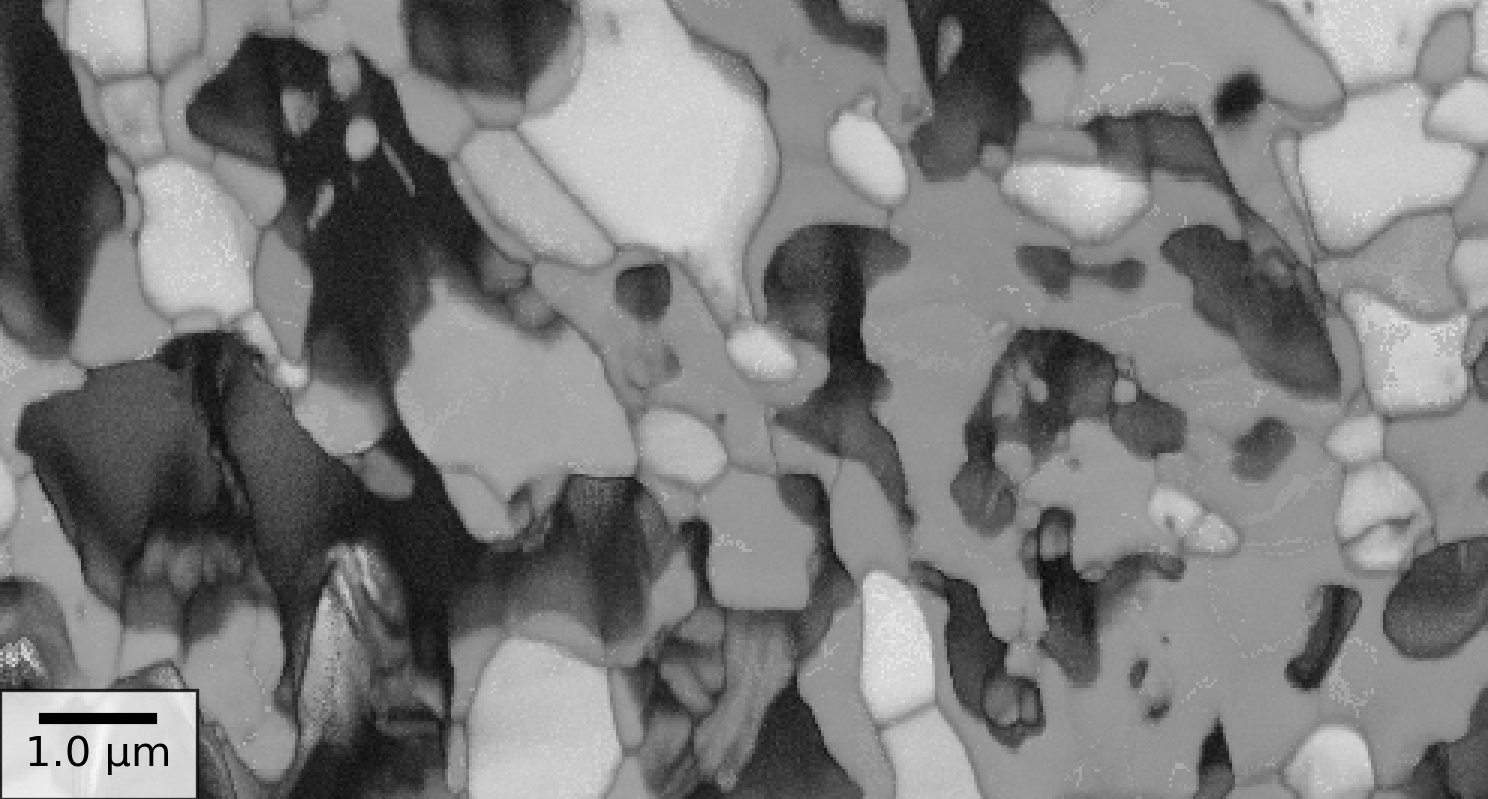}
  \caption{Real space band contrast map.}
  \label{fig:band_contrast}
\end{figure}

The corresponding band contrast (BC) map (Figure~\ref{fig:band_contrast}) highlights different microstructural features (diffraction quality) between metallic iron and oxide grains.
Regions corresponding to \(\alpha\)-iron grains exhibit higher contrast intensity, while the oxide phases appear darker.
A pronounced reduction in BC intensity is observed at grain boundaries, consistent with the degradation of Kikuchi pattern quality near these regions, allowing BC to indicate microstructural features and correlate them with defects and pores.

\begin{figure}[htp!]
\centering
    \begin{subfigure}{.6\textwidth}
        \centering
        \includegraphics[width=\textwidth]{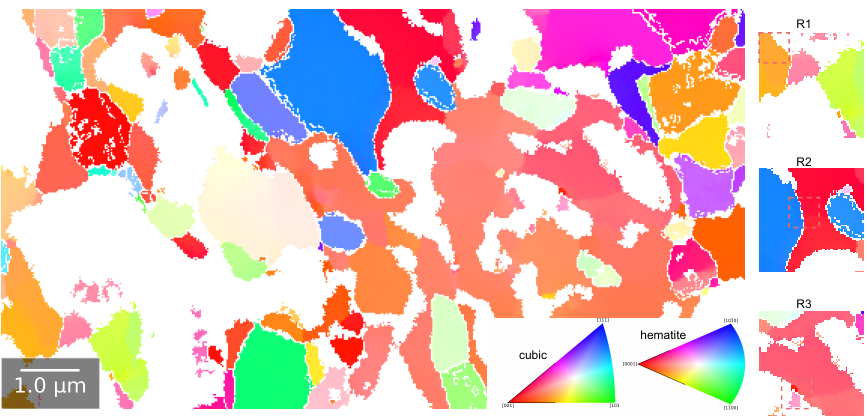}
        \caption{Inverse pole figure along Z-direction.}
        \label{fig:IPF_z}
    \end{subfigure}
    \hfill  
    \begin{subfigure}{0.35\textwidth}
        \centering
        \includegraphics[width=\textwidth]{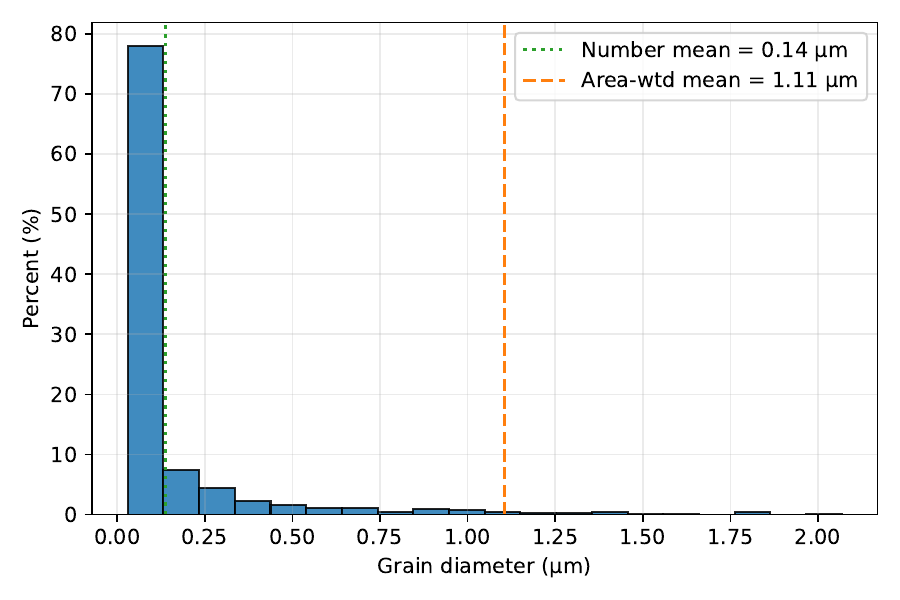}
        \caption{Grain diameter distribution.}
        \label{fig:grain_diameter_distribution}
  \end{subfigure}

  \caption{Inverse pole figure along Z direction of sample and grain diameter distribution
  (a) IPF at Z (R1/R2/R3 (with red dashed boxes highlighting X1/X2/X3 respectively) are corresponding to R regions in phase map); (b) grain diameter distribution}
  \label{fig:IPF_with_grain_diameter}
  
\end{figure}

The inverse pole figure (IPF) map along the \(Z\) direction of the sample is shown in Figure~\ref{fig:IPF_z}.
In this representation, different phases are distinguished by their corresponding crystal symmetries.
While hematite exhibits a hexagonal prism structure, represented with a distinct color key, the other phases—including magnetite, w\"ustite, and \(\alpha\)-iron—belong to the cubic system, hence sharing the same IPF color legend.

The corresponding grain size distribution is shown in Figure~\ref{fig:grain_diameter_distribution}.
It shows the area-weighted mean grain diameter of approximately \(1.1\,\upmu\mathrm{m}\), while the number-average mean grain diameter is \(0.14\,\upmu\mathrm{m}\).
Most grains are distributed below \(0.25\,\upmu\mathrm{m}\), consistent with the small-grain microstructure expected for hydrogen-reduced iron oxides under similar experimental conditions~\cite{ratzkerElucidatingMicrostructureEvolution2025,kimInfluenceMicrostructureAtomicscale2021b}.
The grain size is computed using a misorientation threshold of \(2^{\circ}\) to delineate grain boundaries, which aligns with the general setting for sub-grain structures and ultra-low misorientation measurement used in conventional EBSD-based analysis~\cite{lehtoAdaptiveDomainMisorientation2021}.
Grain boundaries with a misorientation of less than 2$^{\circ}$ are regarded unreliable and thus excluded~\cite{sunDelineatingUltraLowMisorientation2024}.
Significantly, the resulting mean grain size also represents a characteristic length scale relevant to the selection of the ROI window used for the data-driven analysis, as discussed in Section~\ref{subsubsec:roi_selection}.

\begin{figure}[htp!]
  \centering
  \includegraphics[width=.6\textwidth]{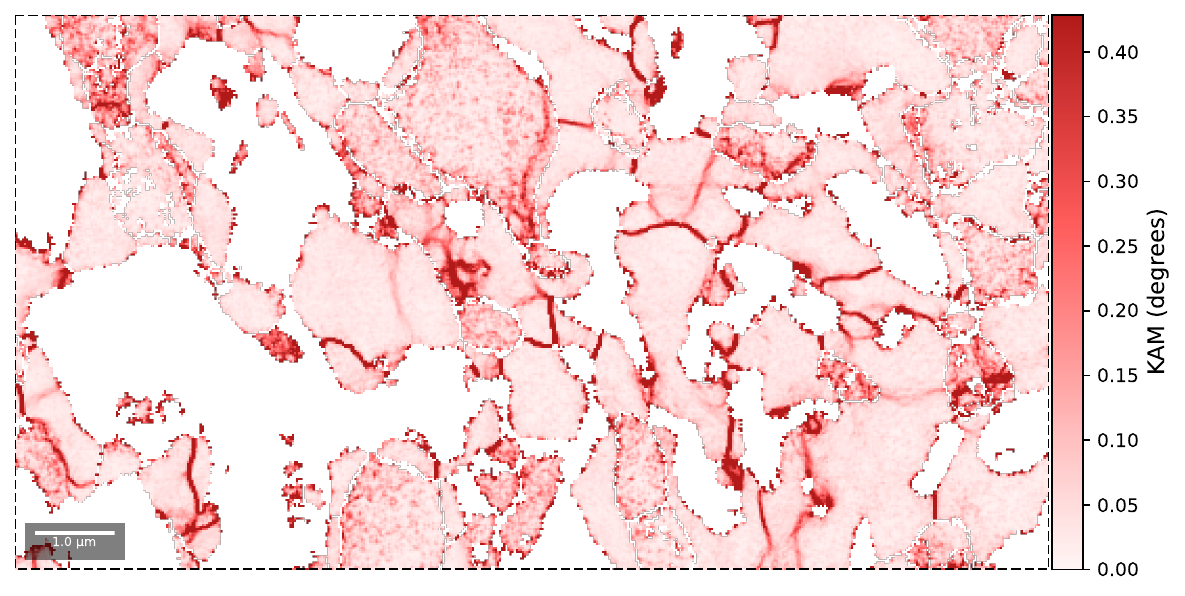}
  \caption{Kernel average misorientation map using 4 nearest neighbors.}
  \label{fig:kam}
\end{figure}

The Kernel Average Misorientation (KAM) map is computed using 4 nearest neighbors to quantify the local lattice curvature and misorientation (Figure~\ref{fig:kam}), in which the value for each pixel is calculated as the the average misorientation between that pixel and its four nearest neighbors.
In EBSD analysis, misorientations below \(15^{\circ}\) are typically interpreted as low-angle grain boundaries (LAGBs) or subgrain boundaries, reflecting the presence of an array of geometrically necessary dislocations rather than fully developed high-angle grain boundaries~\cite{deGraef2012}.  
More specifically, ultra-low misorientations below \(2^{\circ}\) are widely recognized as indicators of subtle crystallographic bending or orientation gradients within individual grains, rather than true grain boundaries~\cite{sunDelineatingUltraLowMisorientation2024}.
In the present dataset, the KAM map reveals that a substantial portion of local misorientation values fall below \(2^{\circ}\).
It could visualize intra-granular orientation variations arising from microstructural evolution and serve as a reference for identifying delicate orientation gradients, which can be sensitively detected using both the PCA–GMM clustering and cNMF analysis as shown in Section~\ref{ssec:data_driven_characterization}.

This analysis serves as a baseline and we will use it along with the data-driven characterization in the following as a reference.

\subsection{Data-driven Characterization}
\label{ssec:data_driven_characterization}

In this section, the data-driven characterization of the same EBSD dataset is presented structured into three parts.  
First, in Section~\ref{sssec:grain_segmentation_boundary_detection}, we apply the unsupervised cNMF workflow to six selected ROIs and compare the resulting grain segmentation and boundary detection with the experimental ground truth obtained from conventional EBSD analysis.  
We show how PCA-GMM clustering and Mahalanobis-distance-based anomaly detection segment the microstructure and highlight grain and phase boundaries, and how these results compare with the cNMF-based spatial weight maps and interface identification.  

Second, Section~\ref{subsubsec:roi_selection} examines the dependence of the cNMF workflow on the chosen ROI size and identifies an ``optimal length scale'' for robust segmentation.  
This characteristic scale is further related to the experimentally determined grain diameter distribution, thereby linking the performance of the data-driven method to an intrinsic microstructural length scale.  

Third, in Section~\ref{sssec:vae_representation}, we employ a variational autoencoder (VAE) to obtain latent representations of the Kikuchi patterns and explore how clustering in latent space separates metallic iron from reduced iron oxides in terms of their spatial distributions.  
Furthermore, we use variance-based ranking and univariate \(F\)-tests, complemented by Pearson, Spearman, and Kendall correlation analysis, to investigate potential relationships between selected latent dimensions and EDS-derived elemental distributions (in particular Fe and O) as well as band contrast, which reflects diffraction pattern quality.

\subsubsection{Grain Segmentation and Boundary detection}
\label{sssec:grain_segmentation_boundary_detection}
\begin{figure*}[htbp]
  \centering

  % ----- (a) Experimental analysis -----
  \begin{subfigure}[t]{0.51\textwidth}
    \centering
    \includegraphics[width=\linewidth]{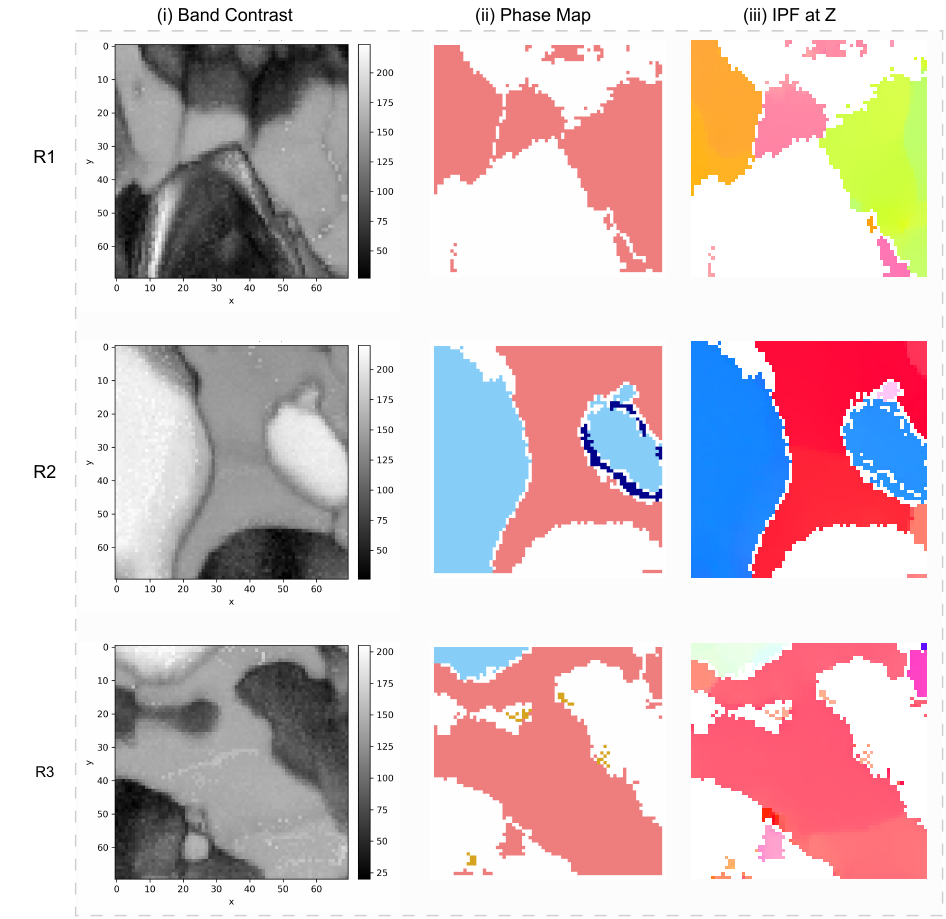}
    \caption{Conventional analysis.}
    \label{fig:r_experimental_characterization}
  \end{subfigure}
  \hfill
  % ----- (b) Data-driven PCA and GMM clustering -----
  \begin{subfigure}[t]{0.47\textwidth}
    \centering
    \includegraphics[width=\linewidth]{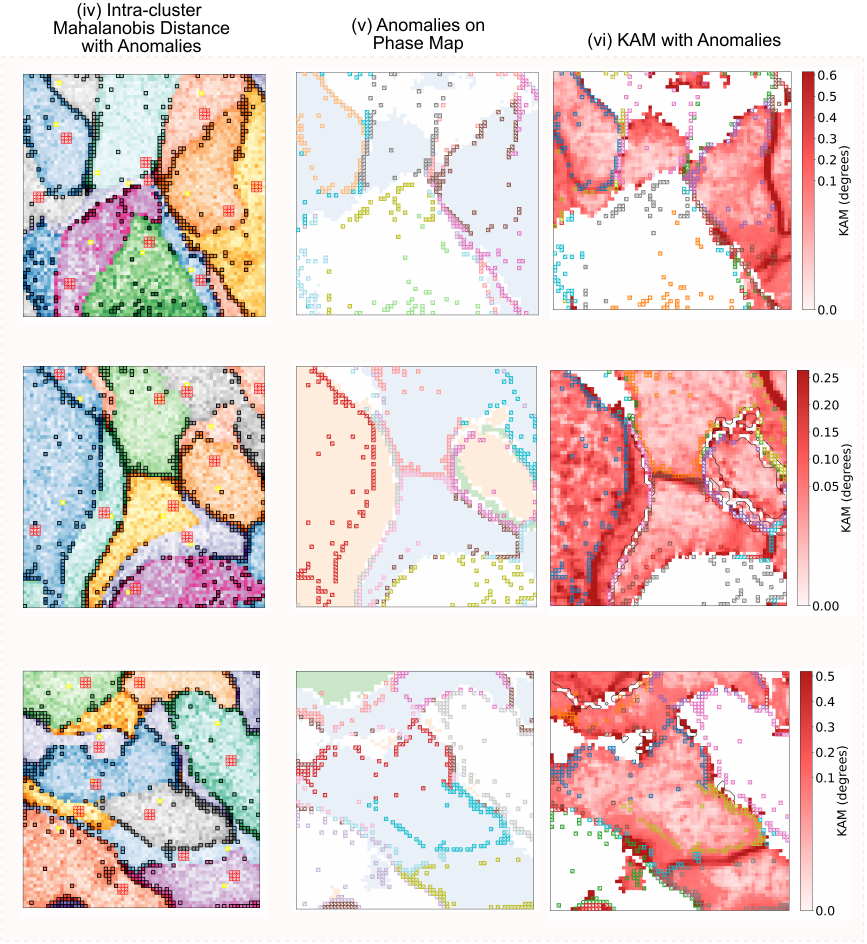}
    \caption{Data-driven PCA and GMM clustering results.}
    \label{fig:r_data-driven_characterization}
  \end{subfigure}

  \caption{
  Comparison between (a) experimental EBSD-based characterization and (b) data-driven PCA and GMM clustering results across the three selected regions of interest (R1, R2, R3). Each ROI consists of six subplots:
  (i) band contrast map, 
  (ii) phase map, 
  (iii) inverse pole figure (IPF) along the \(Z\) direction, 
  (iv) Intra-cluster Mahalanobis distance map with detected anomalies (black boxes), 
  (v) anomalies projected on the corresponding phase map, and 
  (vi) KAM map illustrating the correlation between anomalies and local misorientation. 
  }
  \label{fig:R1R2R3_experimental_vs_datadriven}
\end{figure*}

\begin{figure*}[htbp]
  \centering

  % ----- (a) Experimental analysis -----
  \begin{subfigure}[t]{0.36\textwidth}
    \centering
    \includegraphics[width=\linewidth]{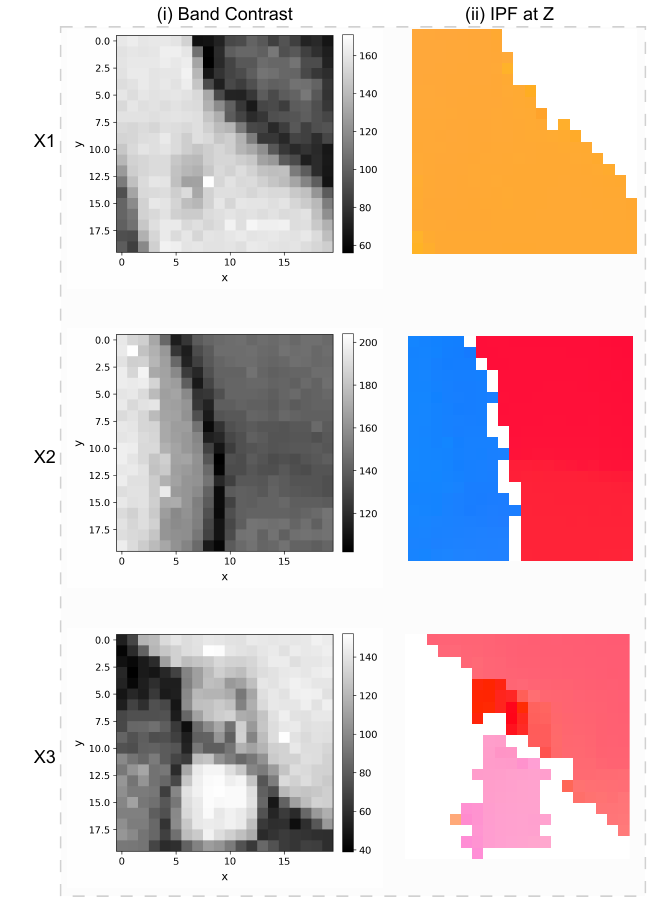}
    \caption{Experimental analysis}
    \label{fig:Experimental_characterization_x}
  \end{subfigure}
  \hfill
  % ----- (b) Data-driven cnmf results -----
  \begin{subfigure}[t]{0.63\textwidth}
    \centering
    \includegraphics[width=\linewidth]{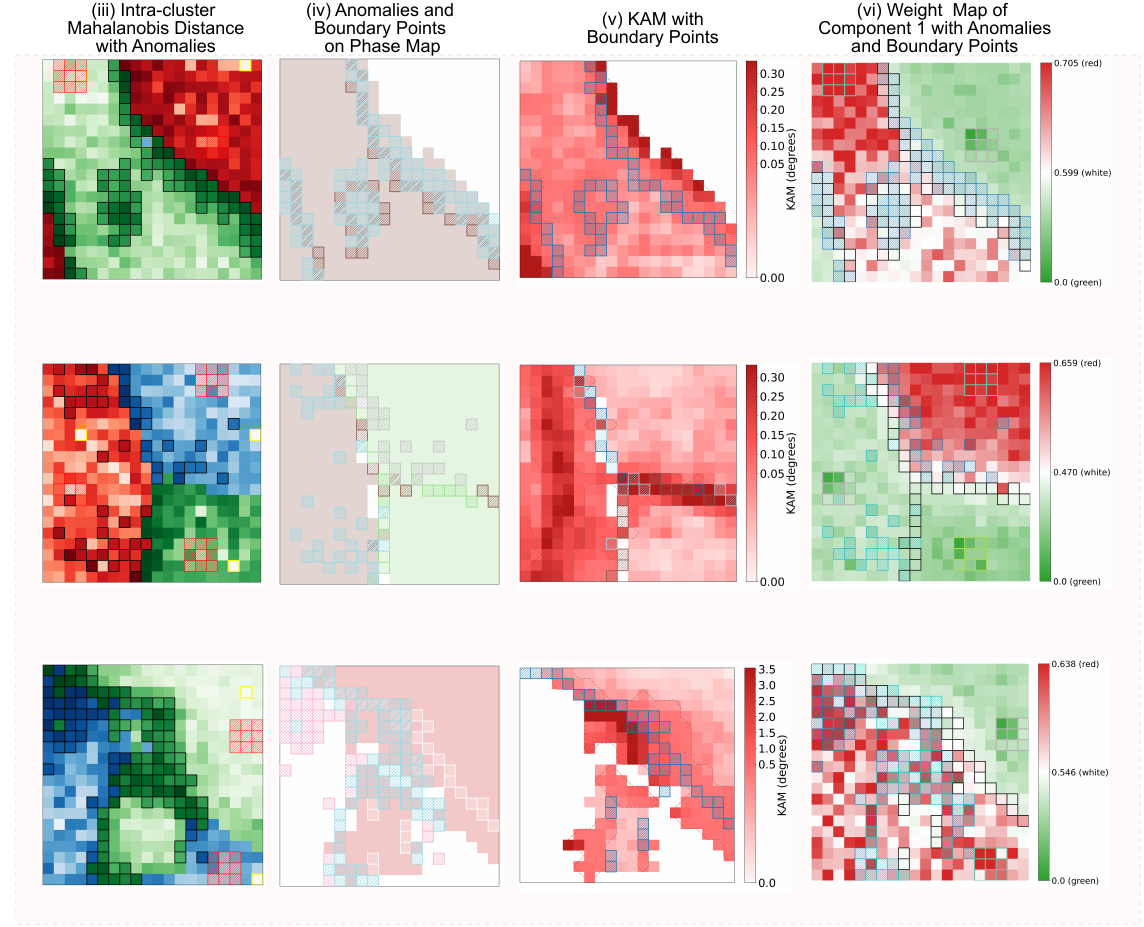}
    \caption{Data-driven PCA–GMM clustering and cNMF analysis.}
    \label{fig:Data-driven_characterization_x}
  \end{subfigure}

  \caption{
  Comparison between (a) experimental EBSD-based characterization and (b) data-driven PCA–GMM clustering and cNMF analysis across three subregions (X1, X2, X3).  
  Each ROI consists of six subplots:  
  (i) band contrast map,   
  (ii) inverse pole figure (IPF) along the \(Z\) direction,  
  (iii) intra-cluster Mahalanobis distance map with detected anomalies (black boxes),  
  (iv) anomalies (hatched box) and boundary points (slashed box) projected on the corresponding phase map, and  
  (v) KAM map illustrating the correlation between boundary points (slashed box) and local misorientation,  
  (vi) the spatial weight map of the first component, in which the intersections of sharp weight transitions, highlighted by black boxes, correspond to the anomalies identified by the PCA–GMM analysis. 
  }
  \label{fig:X1X2X3_experimental_vs_datadriven}
\end{figure*}

The results of both experimental characterization and data-driven analysis covering six representative regions of interest (ROIs: R1-R3, X1-X3) are presented in Figure~\ref{fig:R1R2R3_experimental_vs_datadriven} and Figure~\ref{fig:X1X2X3_experimental_vs_datadriven}.

In both Figures for each ROI, Subfigure (i)–(iii) show the standard EBSD analysis results as reference.
This includes the band contrast map, phase map, and inverse pole figure (IPF) along the \(Z\) direction. 
Subfigure (iv)–(vi) show the results of our unsupervised data-driven approach, where the GMM clustering is performed on the PCA-reduced features.  
The Mahalanobis distance maps (iv) of each cluster reveal intra-cluster variations and highlight statistical anomalies.
Cluster centers are marked with yellow boxes, and the selected \(3\times3\) reference windows used for constraints of cNMF method are indicated by red boxes.  
Points that exceed the 95\,\% confidence interval of the \(\chi^2\) distribution are identified as anomalies and marked with black boxes.
Most of these anomalies are located at the physical boundaries of the corresponding clusters.  
A cross-comparison of these points with the phase maps and Kernel Average Misorientation (KAM) maps shows that most anomalies coincide with phase or grain transition regions (boundaries).

For example, in Region R2, subplot (v) clearly shows that anomalies align closely with both  
(i) the phase boundary between the left orange magnetite grain and the central light-blue \(\alpha\)-iron grain, and (ii) the grain boundary separating the upper and lower parts of this central \(\alpha\)-iron grain, while the left magnetite grain exhibits pronounced intragranular orientation gradients (subgrain structure), which are also captured as anomalies by computing the Mahalanobis distance for each point with respect to its assigned cluster.
Similarly, in Region R3, the high-misorientation line separating the lower-right grain and the central grain, as well as the interface between the grain and adjacent pores, are successfully identified.
The pore-grain interfaces are then detected as anomalies, corresponding to scan points where the signal gradually transitions from noise-like pore patterns to increasingly well-defined Kikuchi bands within the grain interior.

Furthermore, the correlation between anomalies and grain boundaries can be clearly observed in the smaller subregions X1-X3 (Figure~\ref{fig:X1X2X3_experimental_vs_datadriven}).  
In these regions, anomalies are not only associated with edge areas of grains but also appear within grains which we interpret as internal defects.
For example, in region X1 (Figure~\ref{fig:Data-driven_characterization_x} (iii)), several anomalies are located inside the central grain in areas where the band contrast is locally reduced, suggesting partial degradation of Kikuchi pattern quality.
Region X3 also exhibits a spatial aggregation of anomalies along the interface between the primary grain above and an adjacent fine-grained region at the bottom.
This transition zone likely involves grain subdivision or refinement during iron formation.  
Some of these anomalies may correspond to regions affected by noise due to lattice distortion or other crystallographic defects.
This observation is consistent with the interpretation that anomalies represent data points whose feature descriptors, namely the PCA-reduced scores and cNMF-derived component weights, deviate significantly from those of the dominant Kikuchi pattern reference within their respective grains (or clusters).

The cNMF approach also demonstrates a strong capability in segmenting microstructural features at the lower X1-X3 scale, as shown in Figure~\ref{fig:X1X2X3_experimental_vs_datadriven}.
This segmentation is achieved by analyzing the weight contributions of different reference components for each sample, where these reference components are automatically derived from the preceding PCA and GMM clustering steps of the larger ROI.

\begin{figure*}[htbp]
  \centering

  % ----- (a) X1 -----
  \begin{subfigure}[t]{0.3\textwidth}
    \centering
    \includegraphics[width=\linewidth]{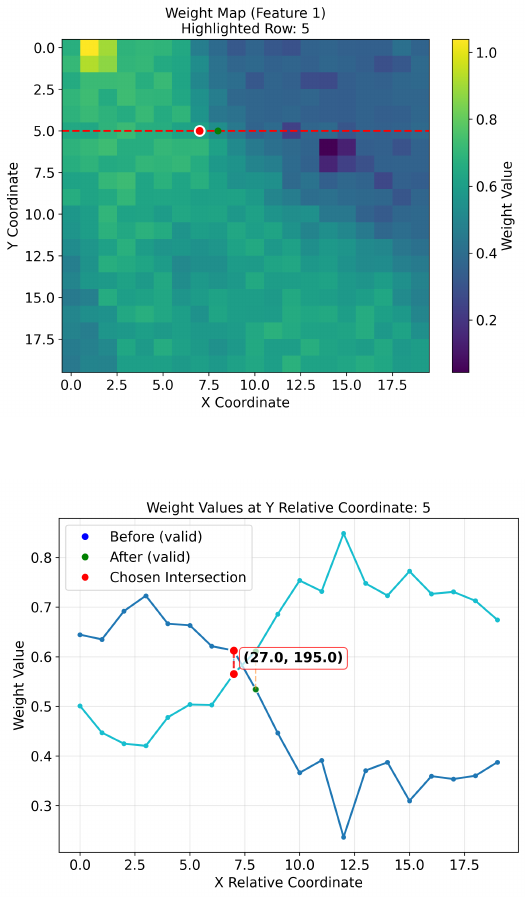}
    \caption{X1}
    \label{fig:x1_weight_map_intersection}
  \end{subfigure}
  \hfill
  % ----- (b) X2 -----
  \begin{subfigure}[t]{0.3\textwidth}
    \centering
    \includegraphics[width=\linewidth]{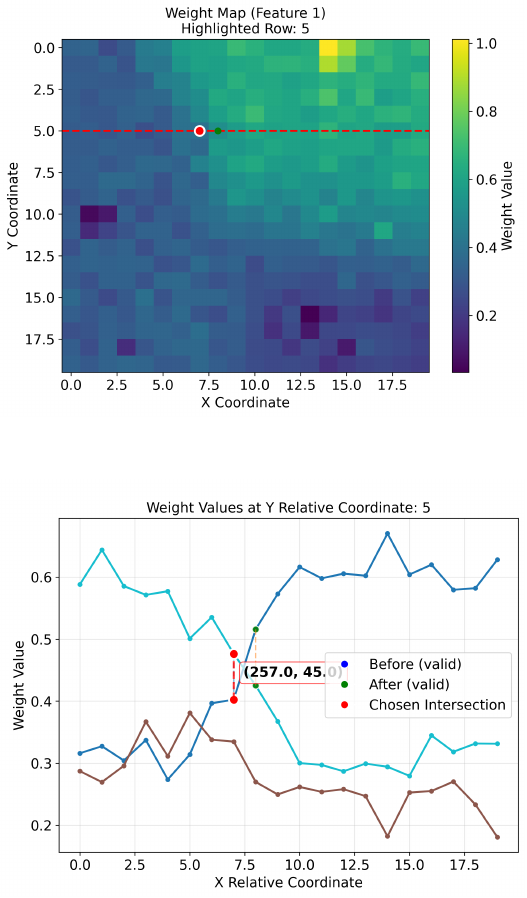}
    \caption{X2}
    \label{fig:x2_weight_map_intersection}
  \end{subfigure}
  \hfill
    % ----- (b) X3 -----
  \begin{subfigure}[t]{0.3\textwidth}
    \centering
    \includegraphics[width=\linewidth]{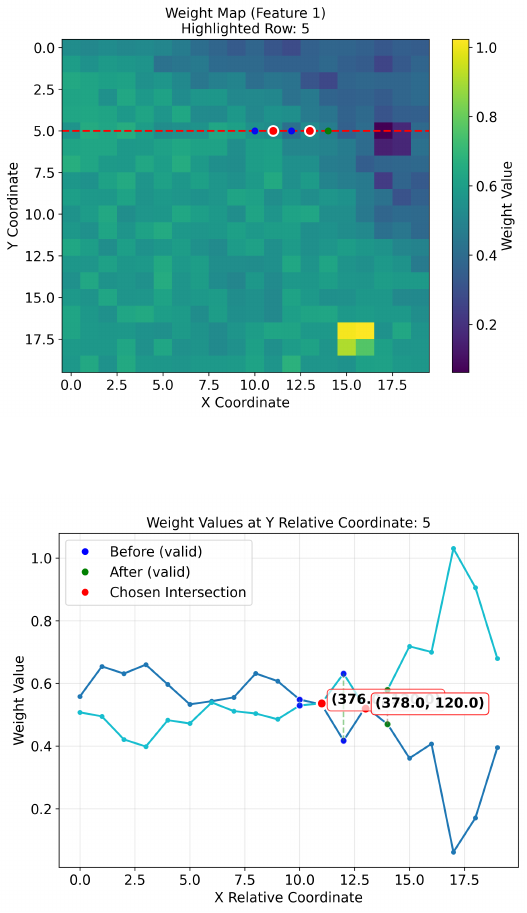}
    \caption{X3}
    \label{fig:x3_weight_map_intersection}
  \end{subfigure}

  \caption{Top row: real space cNMF weight map; bottom row: corresponding weight variations at fixed coordinate Y = 5 in region X1-X3.}
  \label{fig:X1X2X3_variation_curves}
\end{figure*}

Figure~\ref{fig:X1X2X3_variation_curves} (a-c) illustrates how grain interfaces in regions X1-X3 are identified based on the spatial transitions of the cNMF component weights.
For each region, the upper panels show the weight map of the respective cNMF component 1 with the detected interface points extracted at the fixed \(Y=5\) coordinate, while the lower panels present the corresponding one-dimensional weight profiles along that line.
Similarily, the interface points can be also identified along the X axis.

Specifically, for each spatial axis (X or Y), only the two components exhibiting the highest weight magnitudes at each \(x\)-location are considered, and the interface is assigned to the point closest to the upward-downward crossing of these two dominant curves, representing the transition zones between adjacent microstructural constituents in a data-driven manner.
All detected nearest-intersection points are plotted as black-boxed markers in the corresponding spatial weight maps shown in Figure~\ref{fig:Data-driven_characterization_x} (subplots (vi)).  
These results indicate that our automated accurately localizes the grain interfaces and boundary regions, even in cases where the microstructural transitions are subtle or spatially diffuse.

In region X2, at the triple junction, the intersection of the two dominant component weights (whose slopes change in opposite directions) marks the boundary between the neighboring grains.
The weight map in Figure~\ref{fig:Data-driven_characterization_x}(vi) for each component could capture the compositional and structural variations (both phases and orientations) in spatial dimensions, reflecting how the microstructure evolves relative to the selected reference components.
In Figure~\ref{fig:Data-driven_characterization_x}(vi), a clear correlation is observed between the PCA-clustering-based anomalies and the cNMF-detected boundary points.
For the central grain, the reference window is located near its upper-left region, while the weight of this component decreases progressively toward the lower part of the grain, approaching a value of approximately 0.5.  
This reduction in component weight indicates a lower similarity between the local Kikuchi patterns and the reference pattern, and a greater resemblance to the non-indexed region at the upper-right side, corresponding to pores.
Such a transition suggests the presence of lattice defects or local imperfections near the lower grain boundary features that are not discernible through conventional analysis. 

A comparable pattern can be found in Figure~\ref{fig:X1X2X3_experimental_vs_datadriven} X3 (iii), where the second reference component is selected near the lower-right corner of the region, corresponding to the area with the lowest band contrast and evident porosity. 
In the vicinity of the lower-left pores, the corresponding weight map exhibits a less coherent spatial distribution, showing irregular fluctuations rather than a smooth gradient of increasing or decreasing weights.
This irregularity likely reflects the combined effects of sub-grain nucleation occurring within the porous structure and the variation in signal response along the \(Z\)-direction caused by the heterogeneous distribution of pores.

\subsubsection{ROI Selection}
\label{subsubsec:roi_selection}
Our characterization workflow can generally be seen as an extension on top of standard EBSD analysis, in particular to assess small intra-grain variations in misorientation or for the precise detection of boundaries between crystallographicall distinct regions.
Chauniyal \textit{et al.}~\cite{chauniyalEmployingConstrainedNonnegative2024b} showed that the usefulness of cNMF depends on the size of the chosen ROI: a smaller but not too small region is ideal.
There, both the location of the ROI and its size is based on intuition.
Here we exchange the intuition with an unsupervised algorithm automating the choice of the ROI size based on an information criterion.
This leads to an ``optimal length scale'' for a detailed segmentation based on the data itself and the methods used.
The resulting optimal length scale (X-series, cf. Figure~\ref{fig:X1X2X3_experimental_vs_datadriven} and~\ref{fig:X1X2X3_variation_curves}) is physically intuitive since it relates the resolution of the EBSD scan with a microstructural length scale, here the grain size.

\begin{figure}[htp!]
  \centering
  \includegraphics[width=.8\textwidth]{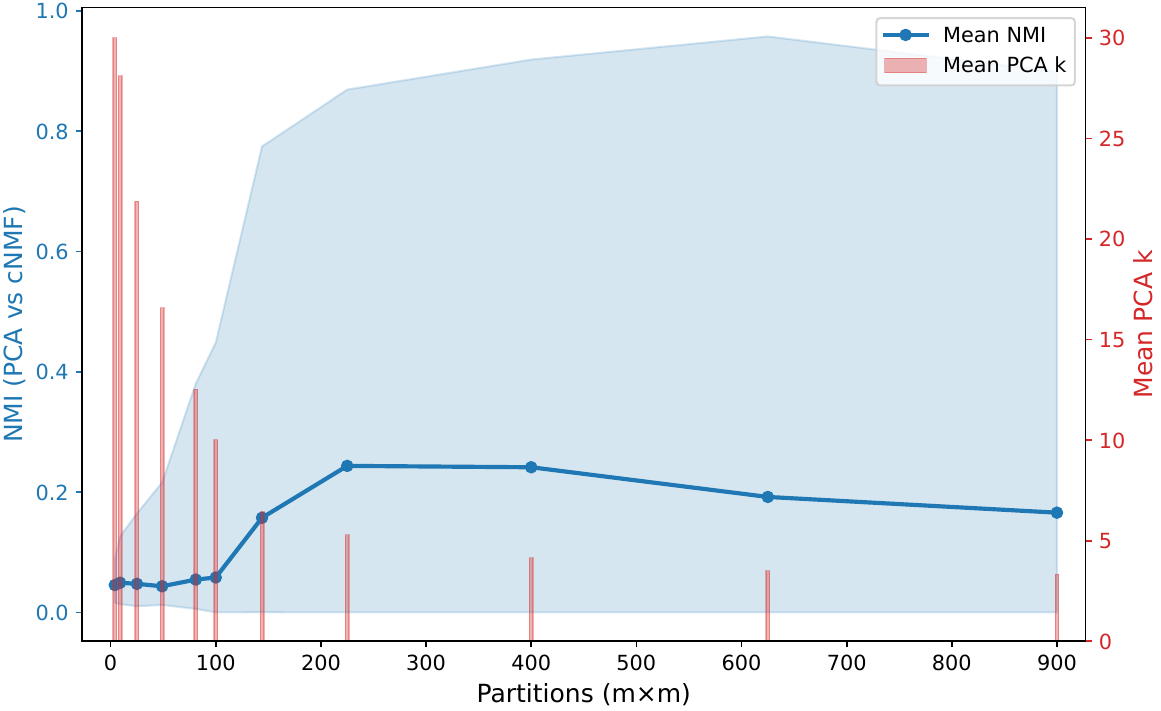}
  \caption{Mean normalized mutual information (NMI) variation and optimal number of clusters in principle component space over partition size.}
  \label{fig:mean_nmi_and_mean_pca_k}
\end{figure}

Figure~\ref{fig:mean_nmi_and_mean_pca_k} shows the Mean Normalized Mutual Information (NMI) average and standard deviation criterion as a function of the partitions (``tiles'') of the whole EBSD scan with a total area of $13 \times 6\,\upmu$m$^2$ ranging from \(2 \times 2\) to \(30 \times 30\) subdivisions.
Averages are obtained by applying the same cNMF workflow, followed by GMM clustering on both the PCA-reduced features and the cNMF weight matrix for each partition.
The clustering consistency between PCA- and cNMF-based representations is then quantified using NMI.

The mutual information between PCA and cNMF representations has its maximum at \(15 \times 15\) partitions (225 partitions, each containing approximately $33 \times 18 = 594$, corresponding to a spatial scale of $0.5$ - $0.9\,\upmu$m).
Furthermore, the maximum NMI values for each partition group exceed 0.8 at \(15 \times 15\) partitions and reach their peak at approximately \(25 \times 25\), corresponding to a characteristic spatial range of $0.3$-$0.9\,\upmu$m.
This range for a \textit{good} ROI size correlates well with an intuitive expectation: the microstructural length scale of grain diameter is $0.14\,\upmu$m (cf. Figure~\ref{fig:grain_diameter_distribution}) and one would expect from the Nyquist Theorem~\cite{Nyquist1928} which states to effectively sample a certain length scale, the sampling length scale must be double -- here ROI size is approx. double the average grain diameter.
Such consistency indicates that the cNMF-based segmentation captures the intrinsic structural hierarchy of the microstructure, where the dominant variations are governed by the representative grain dimensions.
It could be understood when the analyzed region is smaller than a single grain, the local variation within each grain becomes minor, resulting in nearly uniform component weights and weak variation gradients that are easily dominated by noise -- i.e. too little variance in the data leads to a bad signal-noise ratio.

Conversely, as the analyzed region expands, the number of reference components (or constraints) increases, leading to more complex and less distinguishable weight distributions, and the weight curves tend to fluctuate around a similar magnitude, again adding noise. 
This behavior can be interpreted as a gradual loss of explanatory capability of each reference component with increasing distance from its local patterns.
In other words, the ability of cNMF to capture fine structural heterogeneity diminishes at larger scales, likely due to the loss of geometric information during the transformation from raw Kikuchi patterns to their reduced feature representations.

\subsubsection{VAE Representation}
\label{sssec:vae_representation}
The latent representations learned by the VAE potentially encode details of microstructural heterogeneity, because a VAEs can capture complex diffraction information and embed it into a smooth, low-dimensional latent space, cf.~\cite{liuGenerativeArtificialIntelligence2023}.  
A latent space representation of a VAE allows then in principle to identify subtle variations and structural transitions that are not directly discernible through conventional analysis methods~\cite{calvatLearningMetalMicrostructural2025b}.
However, since VAEs are generally considered `unsupervised', the major challenge lies in discerning meaningful information within the possibly hundreds of latent dimensions produced by the encoder and their linkage for to physically interpretable microstructural descriptors.

\begin{figure}[htp!]
  \centering
  \includegraphics[width=\textwidth]{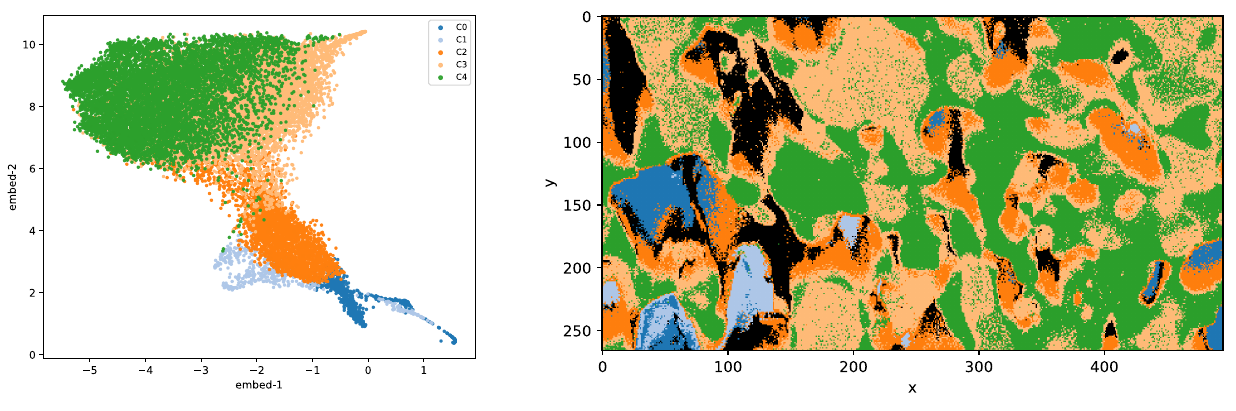}
  \caption{Dimensionality-reduced VAE latent space feature map and real-space coloring using cluster indices by \textit{k}-means clustering.}
  \label{fig:vae_clustering}
\end{figure}

Despite this challenge, we employ VAEs to highlight their potential usefulness in data-driven microstructural characterization.
To explore the structure of the latent space, \textit{k}-means clustering is applied to all 512 latent dimensions and the resulting cluster assignments are visualized using a two-dimensional embedding, as illustrated in Figure~\ref{fig:vae_clustering}.
Note, the dimensionality of the latent space is a user choice, we have not automated this choice.

Since our microstructure also includes pores, we explicitly excluded Kikuchi patterns recorded in pore regions, defined by us using a threshold of band contrast values lower than a normalized value of $0.176$.
We employ this filtering strategy to minimize noise in latent space.
This results in subset of the Kikuchi patterns that excludes Kikuchi patterns associated with a band contrast below the bottom $10\,$\% of all values.
Figure~\ref{fig:vae_clustering} shows these regions in black.
The dimensionality-reduced latent embeddings reveal clear clustering patterns that distinguish metallic iron (green) from the reduced iron oxides (yellow) and other pore regions (blue/light blue and orange).
This observation demonstrates that clustering in latent space very well maps the predominant microstructural characteristics, enabling an unsupervised separation of distinct structural constituents: metallic phases, oxide phases, and pore regions.

To further assess the physical interpretability of the latent features, a feature selection and correlation analysis is performed by combining variance-based pre-screening with univariate \(F\)-tests.  
The elemental spatial distributions obtained from EDS analysis, represented by Fe and O concentration maps, are employed as supervised targets to evaluate the correspondence between VAE latent space features and elemental composition.  
The results indicate that, for both Fe and O, the 1\textsuperscript{st} and the 235\textsuperscript{th} latent dimensions exhibit the strongest correlations with the elemental distributions, respectively.  

In addition, three statistical correlation metrics -- Pearson's linear correlation, Spearman's rank correlation, and Kendall's \(\tau\) -- are computed between each latent feature and the local elemental concentrations.  
Consistently, the 1\textsuperscript{st} and 235\textsuperscript{th} latent dimensions demonstrate the highest correlation coefficients across all three measures for all dataset.

We also assessed the correlations between the VAE latent features and the BC map, which reflects the diffraction pattern quality and local crystallographic perfection/imperfection.
The \(F\)-test identifies the 40\textsuperscript{th} latent dimension as the most strongly correlated with BC.
This finding is further supported by the statistical correlation metrics (Pearson, Spearman, and Kendall), all of which consistently indicated that the 40\textsuperscript{th} latent dimension exhibits the highest correlation with the band contrast distribution.

\begin{figure*}[htbp]
  \centering

  % ----- (a)-----
  \begin{subfigure}[t]{0.3\textwidth}
    \centering
    \includegraphics[width=\linewidth]{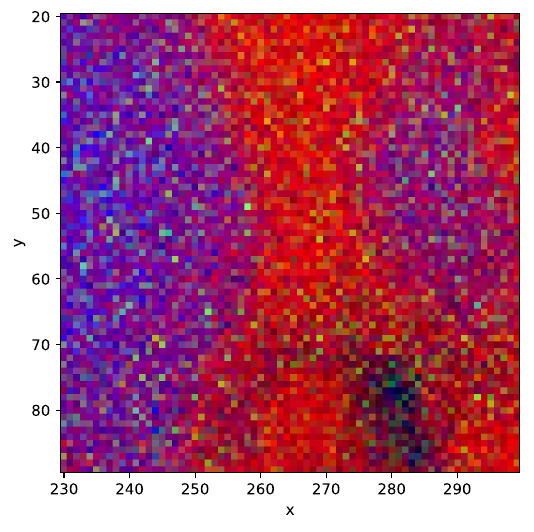}
    \caption{Synthetic RGB elemental map.}
    \label{fig:r2_synthetic_rgb_elemental_map}
  \end{subfigure}
  \hfill
  % ----- (b) -----
  \begin{subfigure}[t]{0.3\textwidth}
    \centering
    \includegraphics[width=\linewidth]{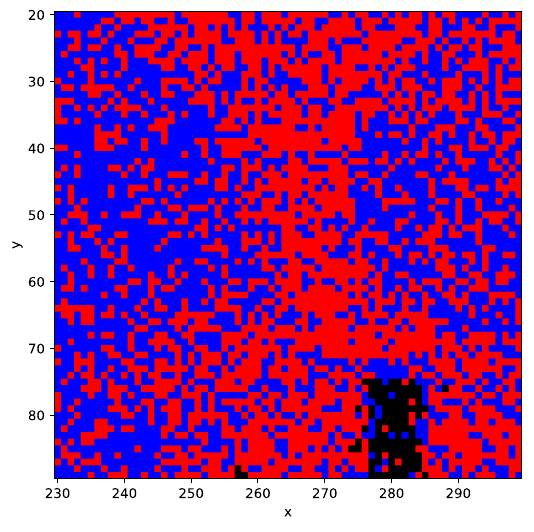}
    \caption{1\textsuperscript{st} latent feature map.}
    \label{fig:r2_the_first_Quantile-band_latent_feature_map}
  \end{subfigure}
  \hfill
    % ----- (b) -----
  \begin{subfigure}[t]{0.3\textwidth}
    \centering
    \includegraphics[width=\linewidth]{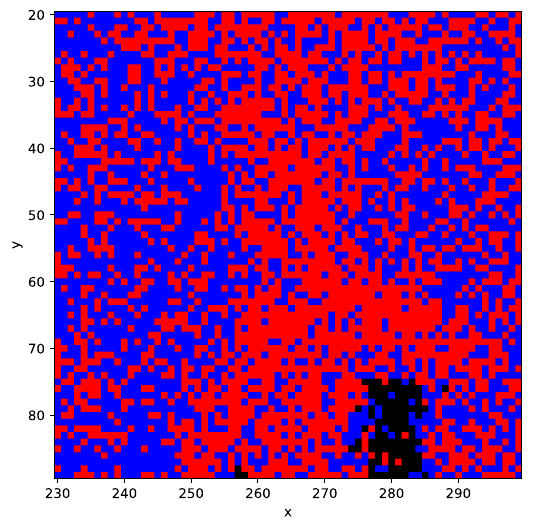}
    \caption{235\textsuperscript{th} latent feature map.}
    \label{fig:r2_235th_Quantile-band_latent_feature_map}
  \end{subfigure}

  \caption{(a) Synthetic RGB element map (Fe-Al-O) of Region R2 (Fe: Red; Al: Green; O: Blue) along with (b) the first and 235\textsuperscript{th} Quantile band latent feature map
  }
  \label{fig:r2_elemental_map_latent_feature}
\end{figure*}

To visualize the spatial correspondence between the identified latent representations and the elemental distributions, the 1\textsuperscript{st} and 235\textsuperscript{th} latent features are mapped using a quantile-banding approach, as shown in Figure~\ref{fig:r2_elemental_map_latent_feature}.  
For each latent dimension, the feature values are divided into two quantile bands based on their distribution.
In other words, the latent values are discretized into quantile-based intervals and the spatial color maps are reconstructed across the region R2 in Figure~\ref{fig:r_experimental_characterization}, thereby highlighting regions with distinct ranges of latent feature intensities.

The resulting spatial maps of the 1\textsuperscript{st} and 235\textsuperscript{th} latent dimensions are compared with a synthetic RGB elemental map, where Fe, Al, and O are represented by the red, green, and blue channels, respectively.  
As shown in Figure~\ref{fig:r2_elemental_map_latent_feature}, the regions with higher latent feature values (in the red quantile band) are predominantly located within Fe-enriched areas, demonstrating a comparatively strong spatial correlation between the latent representations and the underlying elemental distribution.
This consistency suggests that the latent features derived from EBSD-based physical crystallographic information can be associated with the chemical composition captured by EDS mapping, i.e. Kikuchi patterns themselves contain multi-modal structural and chemical information.
We suppose that this can be attributed to the distinct Kikuchi pattern characteristics of \(\alpha\)-Fe and iron oxides, in particular the differences in band locations and intensities.

\section{Discussion}
\label{sec:discussion}
Two statements can be made based on our presented results:
\begin{enumerate}
    \item Data-driven workflows are clearly able to \textit{squeeze out} more of existing raw data and complement conventional analysis.
    \item The interpretation of the newly-created analyses proves challenging.
\end{enumerate}

Both statements could be drawn from previously published studies, e.g.~\cite{McAuliffe2020,chauniyalEmployingConstrainedNonnegative2024b,calvatLearningMetalMicrostructural2025b}.
However, the major difference between these other comparable works and ours is that we arrive at these conclusions through an automated workflow that does not require any user input and, therefore, is not biased but based on information contained in the raw data itself.

We achieve this by a combination of data-driven methods which use the information contained in the data itself rather than (possibly biased) user input.
Many data-driven methods for dimensionality reduction or clustering are supervised or at least semi-supervised and require the setting of hyperparameters.
For PCA, this is the number of components, for cNMF the choice of the constraints (reference patterns), for GMM the number of clusters $k$, for the ROI the resolution, for VAEs the dimensionality of the latent space, etc.
In applications to materials data, these choices are typically made ad-hoc based on domain knowledge and/or the expected results.
Another option is typically to report multiple choices and their comparative analysis.

However, from the point of view of information content, many -- if not all -- of these choices can be made with reproducible criteria that depend only on the data itself.
The best example for this in our study is the choice of the size of the ROI (cf. Section~\ref{subsubsec:roi_selection}): By employing information-theoretical measures such as the NMI, we arrive at an optimal size of the ROI which coincides with a good resolution to characterize the microstructural length scale of the mean grain/phase size (Figure~\ref{fig:grain_diameter_distribution}).

Such an unbiased selection is useful for the subsequent analysis because it provides an optimal choice based on the information contained in the data itself.
Our comparative results demonstrate a clear scale dependence between the cNMF and PCA-GMM clustering application of our dataset of a partially-reduced pellet sample.
The performance of the cNMF-based workflow is particularly sensitive to the chosen ROI size, which must be commensurate with a microstructural length scale such as the grain size to resolve meaningful spatial variations in the component weights.
In contrast, PCA-GMM clustering remains relatively robust over larger ROIs and can reliably separate major microstructural regions; combined with Mahalanobis-distance-based anomaly detection, it also allows the identification of grain and phase boundaries as well as other localized defects.  
However, the direct physical interpretation of PCA-reduced features and cluster centroids in terms of specific microstructural descriptors (e.g., composition or orientation) remains non-trivial.  
By comparison, the spatial cNMF weight maps are, by construction, more directly linked to local mixtures of reference Kikuchi patterns and thus to the spatial variation of microstructural constituents, provided that the ROI size is chosen in a physically meaningful range.

As shown in Figure~\ref{fig:R1R2R3_experimental_vs_datadriven} R2, the PCA-GMM clustering with anomaly detection performs effectively at a larger spatial scale, successfully identifying the subgrain misorientation gradient within the left grain.  
However, when the same approach is applied to smaller regions such as Region~X2 (Figure~\ref{fig:X1X2X3_experimental_vs_datadriven}), its sensitivity decreases, particularly in detecting local orientation gradients near the triple junction.

In contrast, cNMF exhibits superior segmentation accuracy at smaller spatial domains (X1-X3), where the decomposition of local features allows a more precise delineation of grain boundaries and interfacial regions.  
CNMF is inherently sensitive to local structural variations and thus well-suited for analyzing fine-scale microstructural transitions; it is typically less robust for characterizing complex heterogeneities over larger regions.

The PCA-clustering anomaly detection can complement the cNMF workflow by identifying intra-granular imperfections or defects within grains.
For instance, in Figure~\ref{fig:X1X2X3_experimental_vs_datadriven} X3, the PCA-clustering anomalies highlight fine-grain nucleation and local heterogeneity associated with pores.
Complementarily, the cNMF representation captures the overall spatial variation of crystalline microstructures.
While sudden fluctuations in component weight values may occasionally appear in regions of local misorientation or defects, the overall weight distributions generally exhibit a smooth, interpretable trend.  
This makes cNMF particularly effective for resolving interface patterns in smaller regions, such as the triple-junction region in Figure~\ref{fig:X1X2X3_experimental_vs_datadriven} X2, where the intersections of weight transitions correspond closely to the grain and phase boundaries identified experimentally.
These observations suggest that PCA clustering and cNMF offer complementary insights into microstructure characterization, offering pixel-level interpretability that extends beyond conventional indexing-based methods.

Less clear, however, is the usefulness of VAE-based representations.
Although~\cite{calvatLearningMetalMicrostructural2025b} convincingly demonstrate microstructural features beyond standard EBSD analysis, labeling individual latent space dimensions with known crystallographic features remains a challenge.
Admittedly, we also only partially solve this problem by a correlative analysis with EDS measurements at the same scale.
We show that some latent space dimensions correlate well with specific element concentrations but these correlations remain qualitative and depend on the data set.
In other words, there is no guarantee that `latent space dimension 235' \textit{always} correlates with Fe content.
Establishing a direct, physically grounded mapping between individual latent dimensions and well-defined microstructural attributes, such as phase composition, defect density, or crystallographic orientation, remains a significant challenge for now.
For VAE-based analyses, future efforts should focus on refining the encoder-decoder architecture and on the incorporation of physics-informed constraints to improve uniqueness and, therefore, interpretability by mappings to other known chemical or structural modalities.

\section{Conclusion}
We present an automated and robust unsupervised data-driven workflow for grain and phase segmentation as well as boundary detection using a complex microstructure of a partially-reduced iron oxide pellet, integrating principle component-based (PCA) clustering with constrained non-negative matrix factorization (cNMF). 
Our workflow enables fully automatic grain and phase segmentation and local identification of reference components without user input, ensuring consistency and reproducibility across different datasets because the hyperparameters which need to be set are directly and automatically derived from the data itself.
By evaluating the local variation of cNMF weights, our method captures microstructural transitions not only between different phases but also between (small mis-) orientations and thereby increase the resolution of structural characterization beyond that of the EBSD scan itself.
In addition, we demonstrate multi-component factorization, in contrast to previous studies where cNMF was only demonstrated in dual-component or single-phase contexts.

PCA-based cluster anomaly detection, namely out-of-distribution data points, complements cNMF by revealing intra-granular defects and subtle heterogeneities that are hardly accessible through conventional analysis techniques.
This combination of approaches provides an unbiased and physically interpretable means of distinguishing microstructural features, capturing both phase and orientation contrasts simultaneously and accurately.

Furthermore, the data-driven investigation of characterization scales indicates the existence of an optimal length scale for the cNMF workflow.
NMI analysis between PCA- and cNMF-based clustering reveals that the most consistent segmentation results are achieved when the ROI size is approximately twice the mean grain diameter.
This suggests that the cNMF approach is most effective at mesoscopic scales-large enough to capture representative structural variations, yet sufficiently localized to preserve spatial coherence.  
However, the optimal scale varies depending on the local microstructural complexity and the degree of heterogeneity in the analyzed region.

VAE-based analysis remains challenging.
We show that some latent space features correlate with other modalities such as the elemental concentration.
However, these correlations remain quantitative and future VAE-based approaches should focus on physical interpretability.

\medskip
\textbf{Data and code availability} \par 
The source code used is available through a repository here \href{https://gitlab.ruhr-uni-bochum.de/icams-mids/multimodal\_ebsd\_microstructure}{\texttt{https://gitlab.ruhr-uni-bochum.de/icams-mids/multimodal\_ebsd\_microstructure}}.
A selection of the raw Kickuhi patterns is available on Zenodo \href{https://doi.org/10.5281/zenodo.17748462}{10.5281/zenodo.17748462}.
We limited the number of Kikuchi patterns due to dataset size restriction of Zenodo.
The full dataset is available upon request.

% Acknowledgements
\medskip
\textbf{Acknowledgements} \par
All authors thank Özge Özgün and Dierk Raabe (MPISusMat, D\"usseldorf, Germany) for the provision of partially-reduced sample.
QZ and MS greatfully acknowledge funding through a Deutscher Akademischer Austauschdienst e.V. (DAAD) scholarship No. 91925378 in the context of the International Max Planck Research School for Sustainable Metallurgy - from Fundamentals to Engineering Materials (IMPRS SusMat).
MS acknowledges funding by the Deutsche Forschungsgemeinschaft (DFG, German Research Foundation) for CRC1625 -- A05, project number 506711657.

% References
\medskip

% Use the following code if you wish to generate your bibliography with BibTeX;
% replace the string "MSP-template" below with the name(s) of
% the BibTeX data base(s) you want to use.
% The resulting bibliography-output (the content of the .bbl file)
% must be pasted back into this file before submission.
% Please also include your BibTeX data base file(s) in your submission
% so that we can re-run BibTeX if necessary.
%
\bibliographystyle{MSP}
\bibliography{references}

% \textbf{References}\\

\end{document}